\documentclass[12pt]{ws-mml}

\usepackage{float}
\usepackage{amstext}
\usepackage{graphicx}
\usepackage{url}
\usepackage[unicode=true]{hyperref}

\usepackage{color}

\usepackage{verbatim}

\providecommand{\tabularnewline}{\\}

\makeatother

\begin{document}

\title{Dynamical regularities of US equities opening and closing auctions}

\author{Damien Challet}
\address{Math\'ematiques et Informatique pour la Complexit\'e et les Syst\`emes, CentraleSup\'elec, Universit\'e Paris Saclay, 3 rue Joliot-Curie, 91192 Gif-sur-Yvette, France}
\address{Encelade Capital SA, Innovation Park Building C, EPFL, 1015 Lausanne, Switzerland}

\author{Nikita Gourianov}
\address{Physics Department, Oxford University, Clarendon Laboratory, Parks Road, Oxford, OX1 3PU, United Kingdom}
\maketitle

\begin{abstract}
{\color{black}We first investigate the evolution of opening and closing auctions volumes of US equities along the years. We then report dynamical properties of pre-auction periods: the indicative match price is strongly mean-reverting because the imbalance is; the final auction price reacts to a single auction order placement or cancellation in markedly different ways in the opening and closing auctions when computed conditionally on imbalance improving or worsening events; the indicative price reverts towards the mid price of the regular limit order book but is not especially bound to the spread.}
\end{abstract}
\vspace{2ex} 


\centerline{\small {\bf keywords:} auctions; US equities; linear response; imbalance; liquidity}

\section{Introduction}

Many equity exchanges use auctions to define meaningful opening and closing prices
associated with substantial liquidity and to decrease price volatility
near opening and closing times. Despite their practical importance
and the fact that the relative volume traded at the opening and closing
has been steadily growing, as show below, these auctions have not
attracted much work recently, as the attention of the community has
focused on price discovery issues and intraday dynamics. A central
question in the literature is for example the usefulness of auctions. One criterion
is that of market quality, defined for example as the volatility of
the price and bid-ask spreads just after the opening auction or just before the closing auction. Generally, auctions improve market quality (see e.g. \cite{pagano2003closing,chelley2008market,pagano2013call}).

Our focus is rather on the auctions themselves and particularly on
the pre-auction periods. Papers closely related to ours are only a handful.
\cite{gu2008empirical} find a power-law tailed density of the
position order placement on both sides of reference prices in the
Shenzhen Stock Exchange; \cite{kissell2011us} report the average
fraction of the auction volumes with respect to the total daily volume
and their dependence on special days (month end, quarter end, etc.)
and on the capitalization of the assets of US equities. \cite{gu2010empirical}
report that the average density of the auction
limit order book on both sides of the final auction price is well fitted by an exponential distribution and compute the persistence
fluctuation properties of the order size in the Shenzhen Stock Exchange.
More recently, and more in line with our paper, \cite{boussetta2016role}
study the French Stock exchange and find that different kind of market
participants enter the pre-auction periods at markedly different times,
the slow brokers acting first, while high-frequency traders tend to
be active nearer the end of auctions. In the same vein, \cite{bellia2016low}
show how and when low-latency traders (identified as high frequency
traders) add or remove liquidity in the pre-opening
auction of the Tokyo Stock Exchange.  {\color{black}\cite{lehalle2018market} devote part of a chapter to auctions in a spirit close to ours, in particular regarding typical daily activity patterns. Finally, \cite{challet2018strategic} shows that in Paris Stock Exchange, the antagonistic effects of accelerating event rate near the auction time and the decrease of the typical indicative price volatility cannot fully explain the observed diffusion properties of the indicative price, which is likely due to strategic behavior.}

This paper is organized as follows: we first determine the distribution
of the volumes of matched orders at both auction ending times. We then find
rules of thumb to estimate opening and closing volumes from daily
data. Finally, we examine the dynamics of the indicative price, imbalance,
and matched volume. We show in particular that the reaction of the
final auction price to the placement or cancellation of an order is
markedly different in the opening and closing auctions, and that the
indicative auction prices are mostly under-diffusive, especially during
the pre-closing auction period. {\color{black}Finally, we relate the dynamics of the indicative price to that of the limit order book: although the mid price does attract the indicative price, the latter fluctuates rather much and is not specially bound to be in the spread, which, we argue, is due to the sparseness of the auction order book.}

\section{Data }

Each exchange follows its own auction rules and pre-auction information
dissemination procedure. The basic principle of all auctions however
is the same: traders may submit market orders for the auction (buy/sell
a given volume at any price), or limit orders (buy/sell a given quantity
at a given price) which may be valid either for the auction only or
stay in the open-market order book after the auction if they are not matched.
Finally, the matching price is fixed so as to maximize the matched
volume{\color{black}; if such price is not unique, the one closest to the previous close price is selected.} 

{\color{black}What kind of orders and when they may be sent, however, significantly differs between NYSE Arca\footnote{\url{https://www.nyse.com/publicdocs/nyse/markets/nyse-arca/NYSE_Arca_Auctions_Brochure.pdf}, accessed 2018-09-21}} {\color{black}(henceforth
shortened to ARCA), NYSE\footnote{\url{https://www.nyse.com/publicdocs/nyse/markets/nyse/NYSE_Opening_and_Closing_Auctions_Fact_Sheet.pdf.}, accessed 2018-09-21.} and NASDAQ\footnote{\url{https://www.nasdaqtrader.com/content/TechnicalSupport/UserGuides/TradingProducts/crosses/openclosequickguide.pdf}, accessed 2018-09-21.}. }  {\color{black} For example, during the closing auction on ARCA, limit and market auction orders (known as LOC and MOC, respectively) may be sent until one minute before the auction time, then during the last  minute, only orders reducing the imbalance are accepted. For NASDAQ, the cut-off time for LOC/MOC orders has been reduced to 5 minutes in October 2017. Cut-off time is 10 minutes before the closing auction for NYSE, which adds discretionary quotes (D-quotes) that may override imbalance-reducing orders; they are added five minutes before the auction time, may be submitted or modified up to 10 seconds before auction time, and be cancelled at any time. While these differences have no direct influence on the measure of matched volumes, they are likely to change the dynamic properties near the auction times. Because only ARCA continuously disseminates information, we cannot document these differences.}

We use three data from three different sources:
\begin{enumerate}
\item Individual matched order sizes. For assets traded on ARCA, our Thomson-Reuters Tick History data set includes
an order-by-order breakdown of auction volumes from 2009-02-10 to
2014-07-01. To fix notations, $v_{\alpha,k,d}^{x}$ is the volume
(in shares) of the $k$-th matched order for asset $\alpha$ on day
$d$ for auction $x\in\{\text{open,close\}}$. This dataset contains
16,165,407 orders.
\item Daily matched volume for each auction, for the three exchanges, gathered
from the proper flag of Thomson-Reuters tick-by-tick trades data.
We denote by $V_{\alpha,d}^{x}$ the total matched volume of auction
$x$ of security $\alpha$ on day $d$. By definition, $V_{\alpha,d}^{x}=\sum_{k=1}^{N_{\alpha}^{x}(d)}v_{\alpha,k,d}^{x}$
where $N_{\alpha}^{x}(d)$ is the number of matched orders at the
$x$ auction of asset $\alpha$ on day $d$. We shall also denote
the auction price by $p_{\alpha,d}^{x}$ This data set encompasses
4712 US assets from 2010-01-01 to 2016-11-30, which amounts to 3,125,786
open and close auctions (each), among which 179,669 (6\%) for ARCA,
446,396 (14\%) for NASDAQ and 2,499,721 for NYSE (80\%).
\item Finally, we gathered the real-time information disseminated by the
three exchanges from 2016-09-28 to 2018-01-12 for 1076 assets. This
information consists in the indicative price $\pi_{\alpha,d}^{x}(t)$,
the current matched volume at that price $W_{\alpha,d}^{x}(t)$, and
the current imbalance $I_{\alpha,d}^{x}(t)$, defined as the sum of
the market order imbalance and of the limit order imbalance at the
indicative price. While NYSE and NASDAQ only publish a few snapshots
of these quantities per auction, ARCA disseminates up to a few thousand updates for each auction on each day. The ARCA dataset contains 47,246,490 updates for 58 tickers. Our data provider has a limit of about 4 updates per second; while this is sufficient in most cases, it becomes insufficient near the auctions times of very liquid assets such as SPY.
\end{enumerate}

\section{Matched volumes}

We first focus on fairly stylized properties of auctions volumes and then turn to a more refined analysis of the pre-auction dynamics: indicative price, volume, activity rate and relationship with the regular limit order book. 

\subsection{Single order properties}

\begin{figure}
\includegraphics[width=0.5\textwidth]{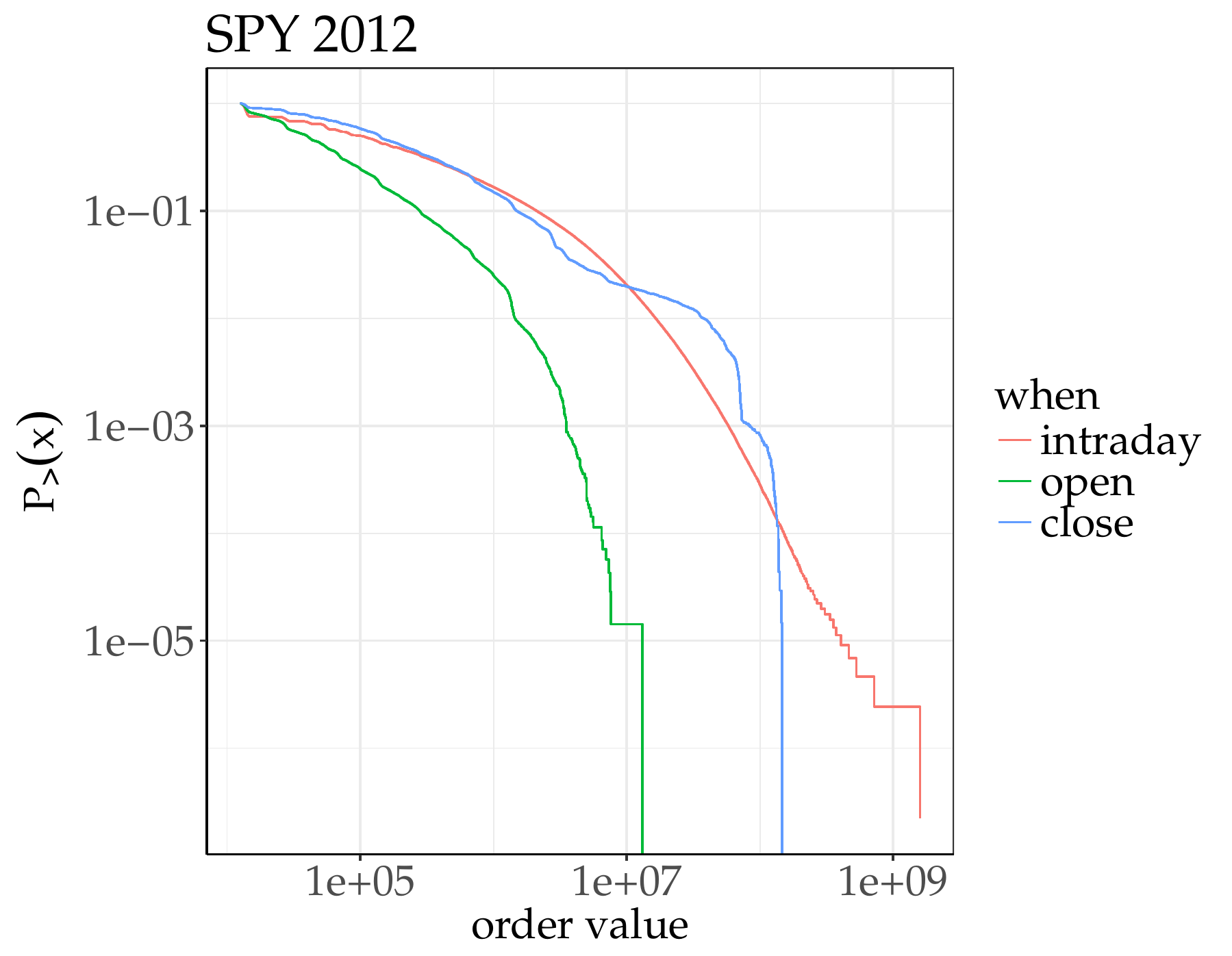}\includegraphics[width=0.5\textwidth]{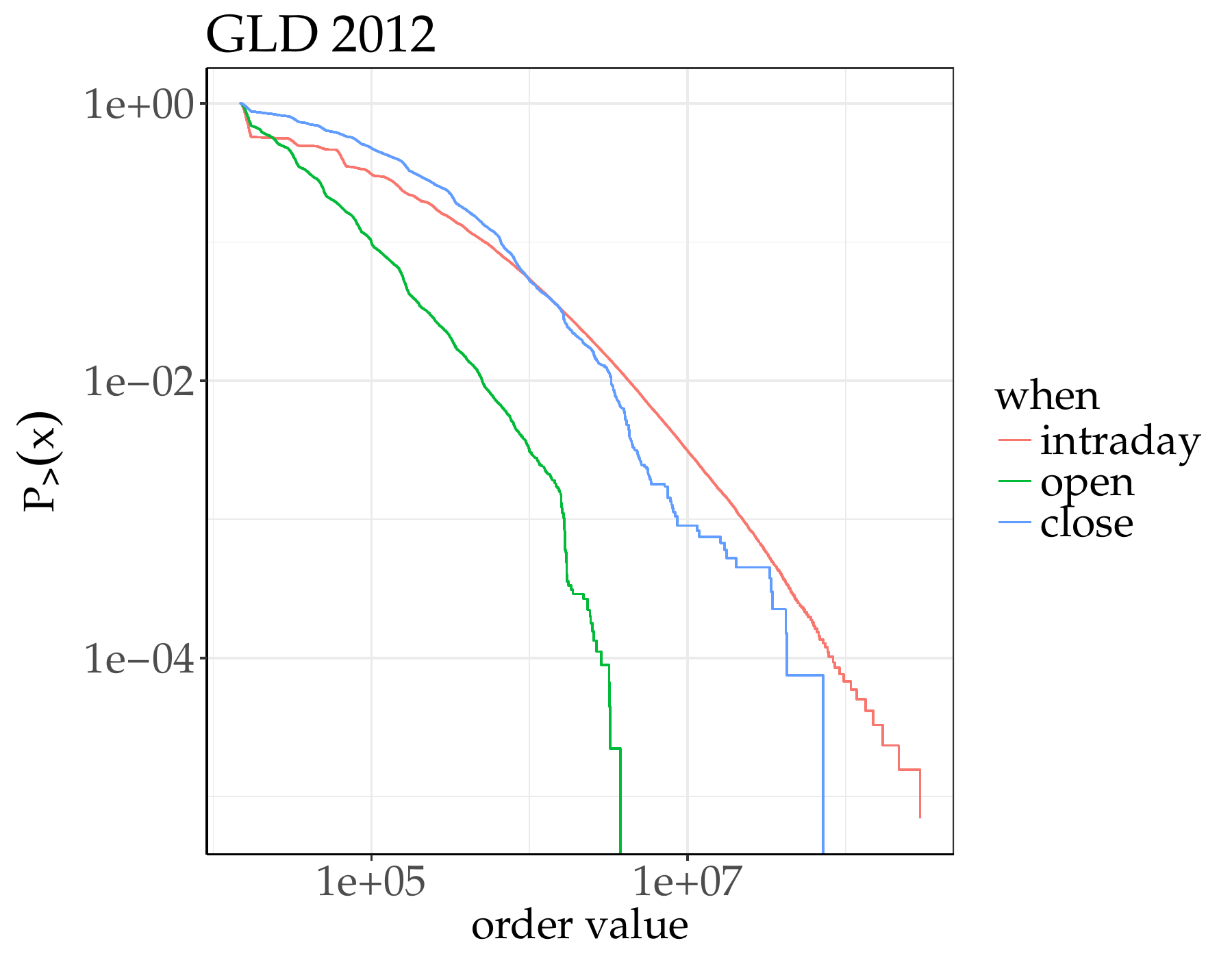}\caption{Reciprocal empirical distribution function of the value of all the
trades of open auctions, close auctions and intraday trading of two
of the most liquid assets traded at ARCA during 2012.\label{fig:ecdf_SPY-2013}}
\end{figure}

Figure \ref{fig:ecdf_SPY-2013} shows the reciprocal empirical distribution
functions of the transaction values of matched orders defined as $p_{\alpha,d}^{x}V_{\alpha,d}^{x}$
at both auctions, together with that of the open-market transaction
values, for SPY and GLD, two of the most liquid assets traded on ARCA
for the whole 2012 year. The distributions of the opening and closing
auctions are clearly much more affected by truncations than that of
the open-market transactions, and the opening auction more than the
closing auction. They reflect, to some extent, the fact that the typical
total volume is larger at the closing auction than at the opening
auction.

\begin{figure}
\includegraphics[width=0.5\textwidth]{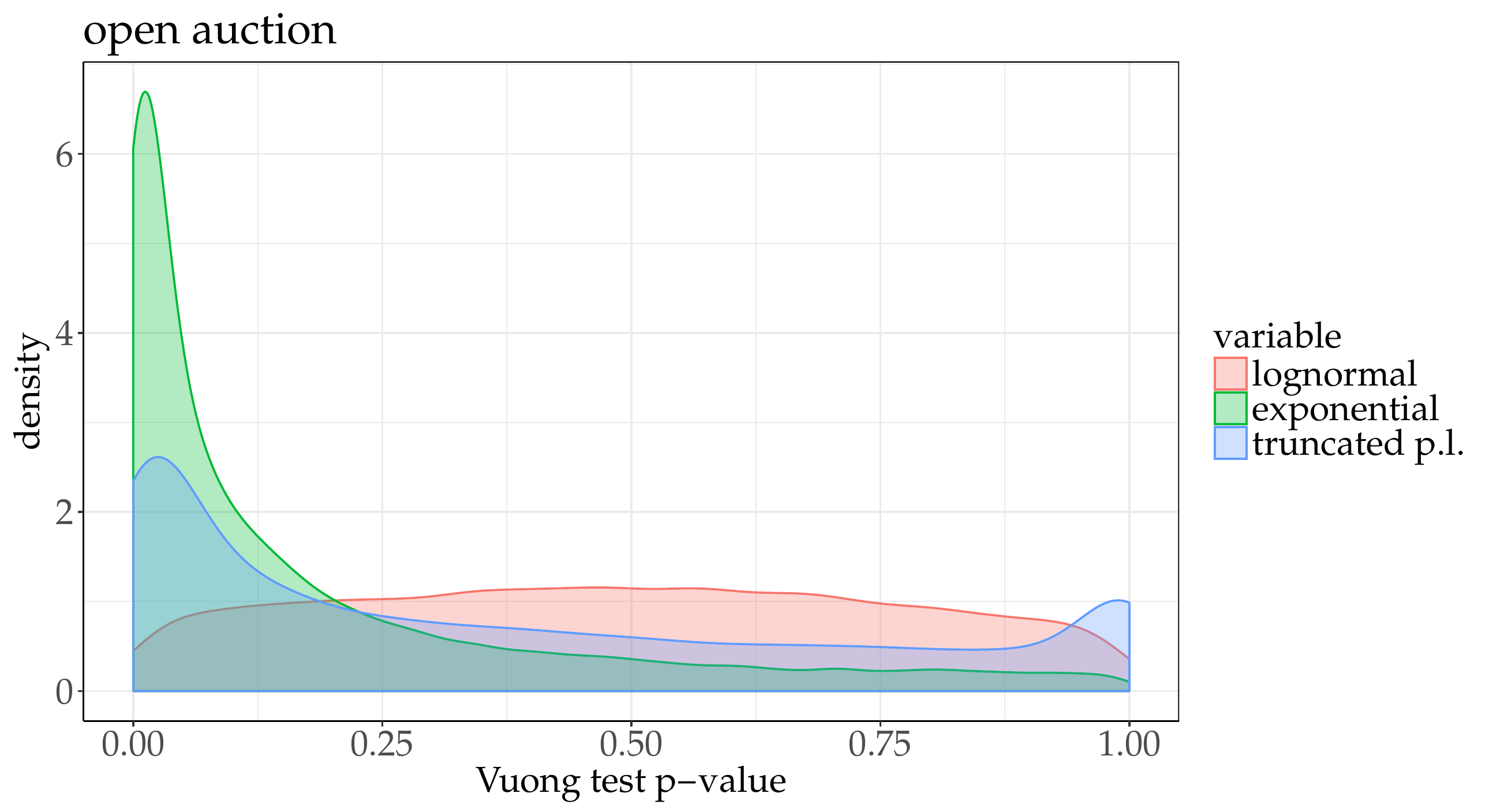}\includegraphics[width=0.5\textwidth]{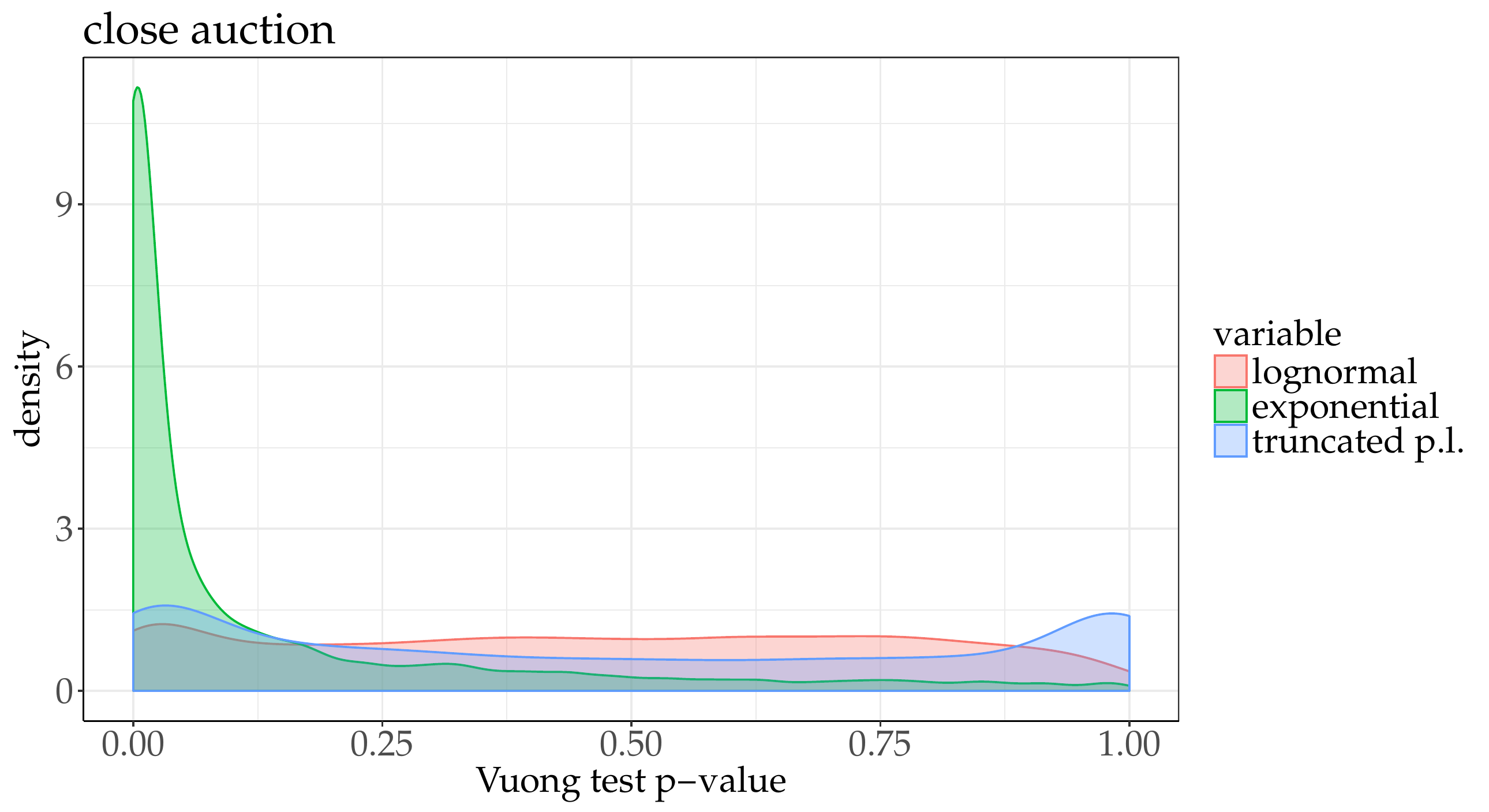}\caption{Tails of the individual matched order volume distribution, for each
day, each asset assessed by the p-values of the Vuong test of a power-law
against log-normal, exponential and truncated power-law.\label{fig:Vuong}}

\end{figure}

This raises the question of the nature of the matched orders distribution
of a single day: does the heavy-tailed nature of $P(p_{\alpha,d}^{x}V_{\alpha,d}^{x})$
come from daily fluctuations or from the distribution of single order
value distributions themselves? Whether the latter have heavy tails
may be assessed by fitting it with an exponential and a log-normal distributions
and using the Vuong closeness test \citep{vuong1989likelihood}. We
perform it for each day and each auction with more than 100 matched
orders, which amount to 29168 auctions. There is  overwhelming evidence
of the presence of heavy tails, corresponding by convention here to
small p-values: about 92.8\% of the auction distributions have a p-value
smaller than 0.01, while only 0.1\% have a p-value larger than 0.99.
Next, we check what distribution may best describe the tails. We use
the method of \cite{clauset2009power,alstott2014powerlaw} to determine
the best starting point of a power-law tail and then investigate how
a fit of that tail with a power-law compares with log-normal, exponential,
and truncated power-law distributions. Figure \ref{fig:Vuong} plots
the density of p-values of a Vuong test between a power-law and
the other candidates, computed for each day, each stock and each auction
with more than 100 matched orders. Expectedly, it confirms the heavy-tailed
nature of the tails. Finally, it shows no real difference between
a power-law and a log-normal distribution, while a truncated power-law
does not bring a substantial improvement on average.

\subsection{Opening, closing and daily volumes}

\begin{figure}
\includegraphics[width=0.5\textwidth]{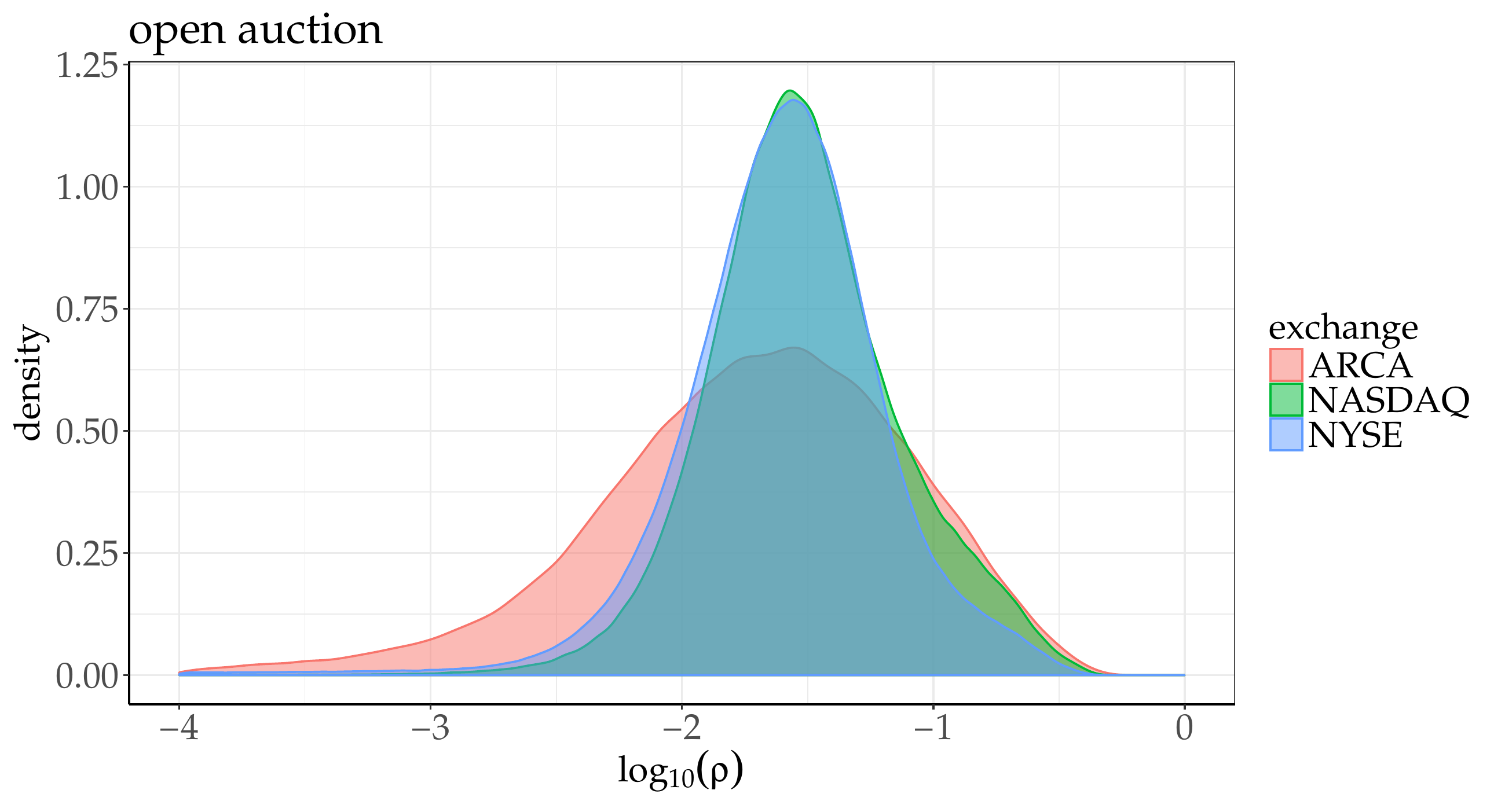}\includegraphics[width=0.5\textwidth]{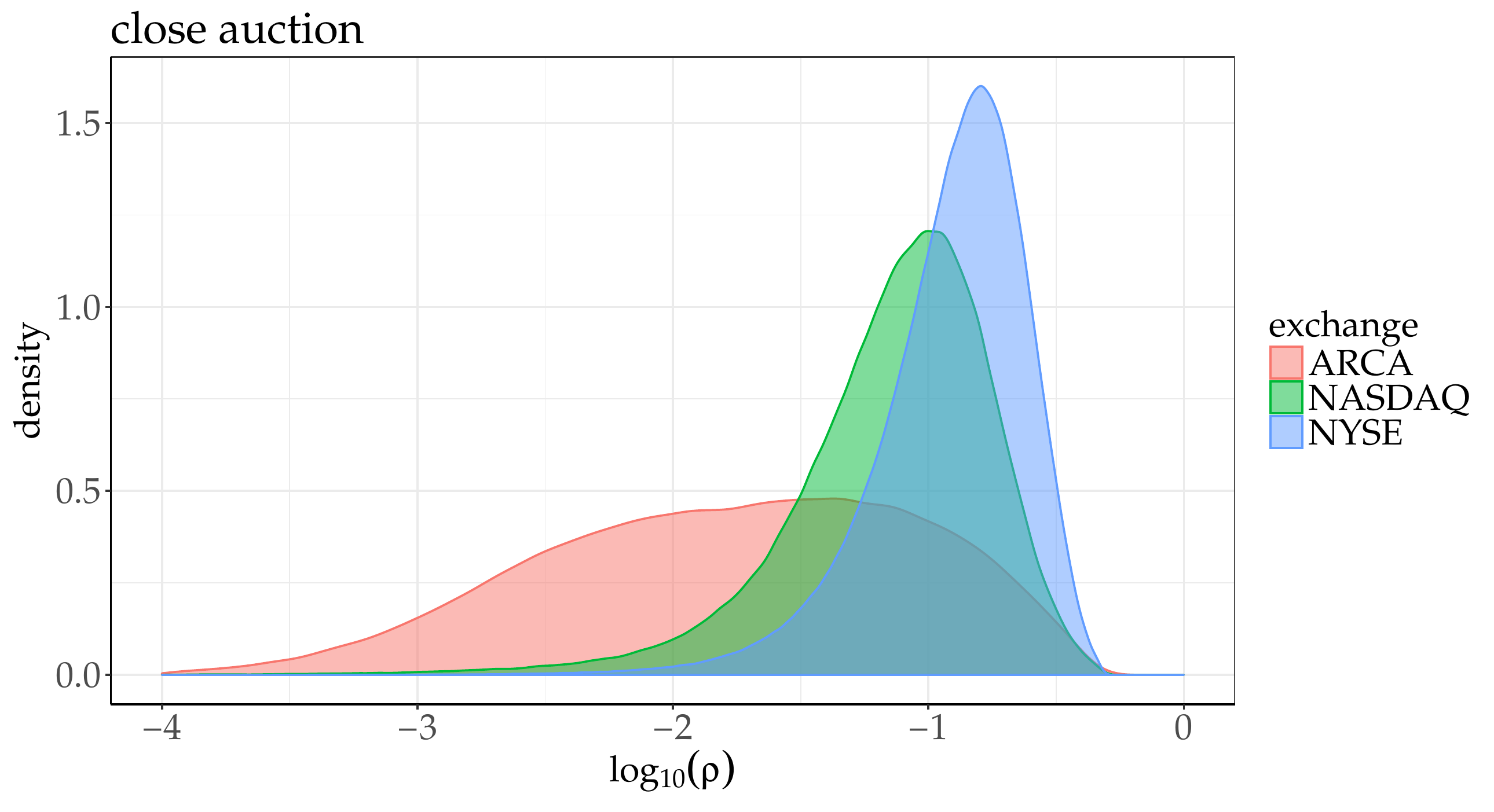}

\includegraphics[width=0.5\textwidth]{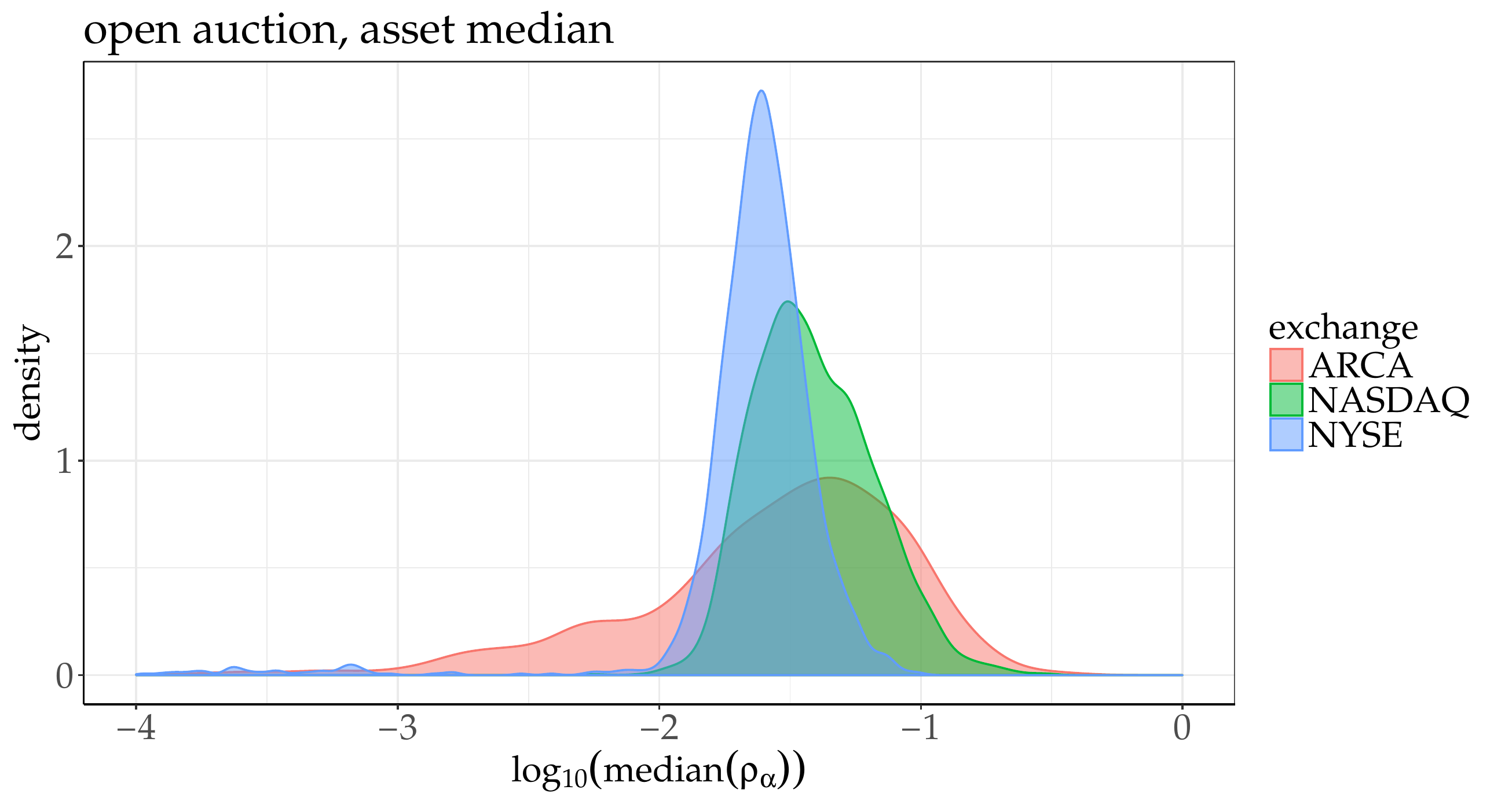}\includegraphics[width=0.5\textwidth]{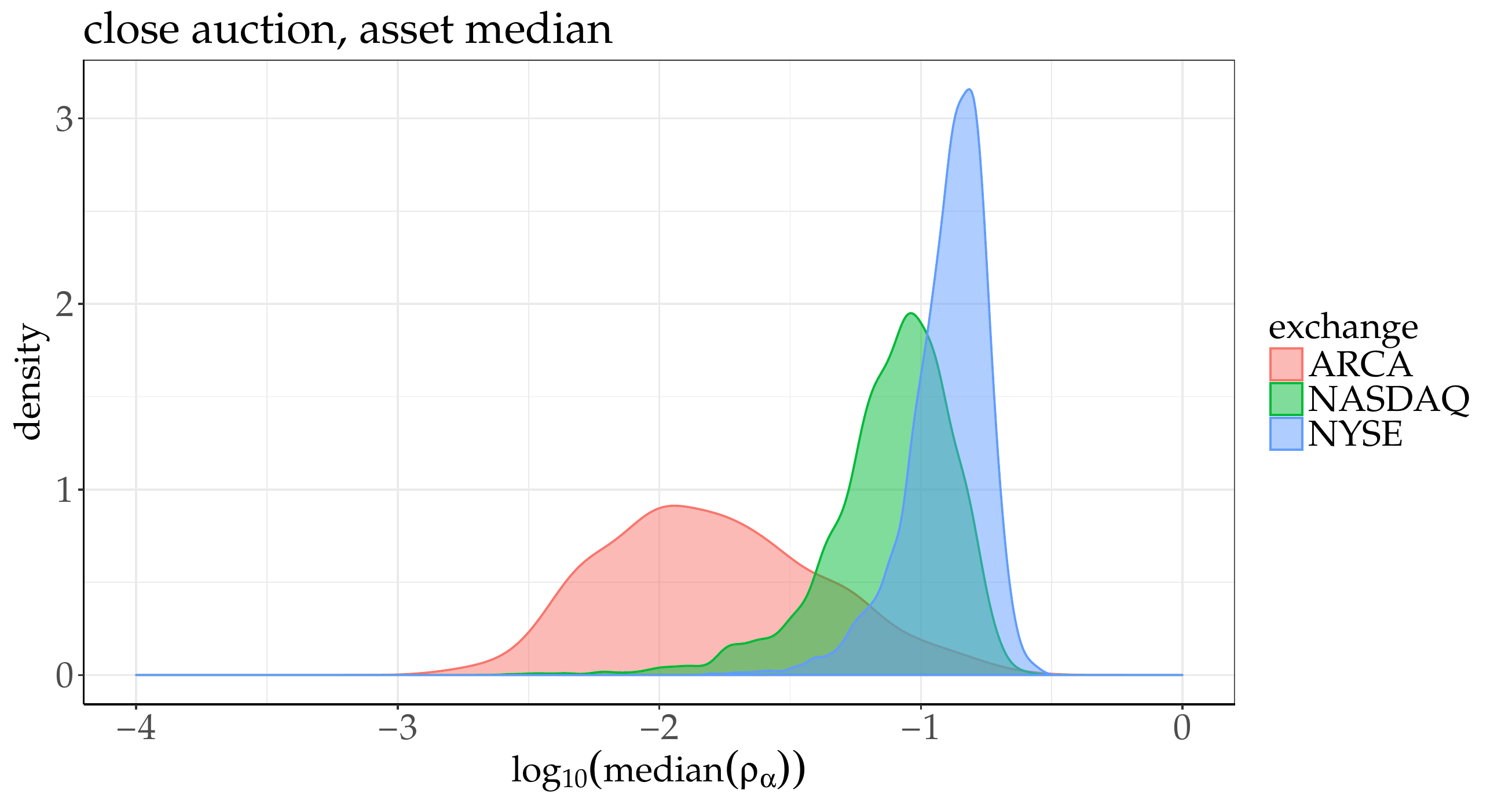}\caption{Upper plots: density of the logarithmic ratio of the auction volumes
to the total daily exchange volume $\rho_{\alpha,d}^{x}=\frac{V_{\alpha,d}^{x}}{V_{\alpha,d}^{\text{total}}}$
for each day and each asset, split by exchange. Lower plotw: density
of the asset median logarithmic ratio, split by exchange. \label{fig:Density-of-median_rho-1}}
\end{figure}

The total daily volume for a given asset is freely available in contrast
to auction volumes. Thus, we endeavor here to find a rule of thumb
to infer the opening and closing auction volumes from daily data only.
Let us focus on the ratios of auction volumes to total daily volume,
denoted by 
\[
\rho_{\alpha,d}^{x}=\frac{V_{\alpha,d}^{x}}{V_{\alpha,d}^{\text{total}}},
\]

where $V_{\alpha,d}^{\text{total}}$ is the total volume exchanged
during day $d$, including both auctions, {\color{black} OTC transactions}, and outside regular trading
hours transactions. We start with averages over the period 2010-01-01
to the end of the dataset. Figure \ref{fig:Density-of-median_rho-1}
plots the densities of $\log_{10}(\rho)$ for the three exchanges,
each day and each asset. The largest difference between the open and
close is observed in the NYSE exchange, possibly because it is not
fully automated. For example, the opening auction ending time is not fixed
and may happen a few minutes after the opening of the exchange itself.
Remarkably, the close of the NYSE has by far the largest percentage
of volume with respect to the other exchanges. NASDAQ has the least
variability between assets and auctions. We also plot in the same
figure the densities of the median logarithmic ratio asset by asset
in order to assess the intrinsic diversity of these ratios between
the assets.\footnote{We use the median in order to avoid accounting for specials days.
See \cite{kissell2011us} for estimates of the typical change associated
to such days.} Once again the variability of the volume ratio between asset of ARCA
is the largest one. The least asset variability for the opening auction
is found in NASDAQ, while NYSE takes the crown for the closing auction
volume fraction, closely followed by NASDAQ. This yields simple rules
of thumb for all the assets: between 2010 and 2016, 12\% of daily
volume is exchanged at the close of NYSE, 3\% the opening of NASDAQ
and 7\% at the close of NASDAQ. The densities of ARCA volume ratios
are too wide to be summarized by simple rules of thumb.

\begin{table}
\small
\begin{center}
\begin{tabular}{|c|c|c|c|}
\hline 
$\log_{10}\rho_{\alpha,d}$ & ARCA & NASDAQ & NYSE\tabularnewline
\hline 
\hline 
open & $-1.72\pm1.24$ & $-1.51\pm0.79$ & $-1.61\pm0.90$\tabularnewline
\hline 
close & $-1.78\pm1.47$ & $-1.15\pm0.82$ & $-0.91\pm0.60$\tabularnewline
\hline 
\end{tabular}

\vspace{4ex}

\begin{tabular}{|c|c|c|c|}
\hline 
$10^{\text{mean}(\log_{10}\rho)}$ & ARCA & NASDAQ & NYSE\tabularnewline
\hline 
\hline 
open & $0.019$ & $0.031$ & $0.024$\tabularnewline
\hline 
close & $0.017$ & $0.071$ & $0.12$\tabularnewline
\hline 
\end{tabular}
\end{center}
\caption{Left table: mean and two standard deviations of the volume ratio per
auction $\log_{10}\rho_{\alpha,d}$, split by exchange. Right table,
typical volume fraction $10^{\text{mean}\log_{10}\rho_{\alpha}}$\label{tab:rho}}
\end{table}

\begin{figure}

\includegraphics[width=0.5\textwidth]{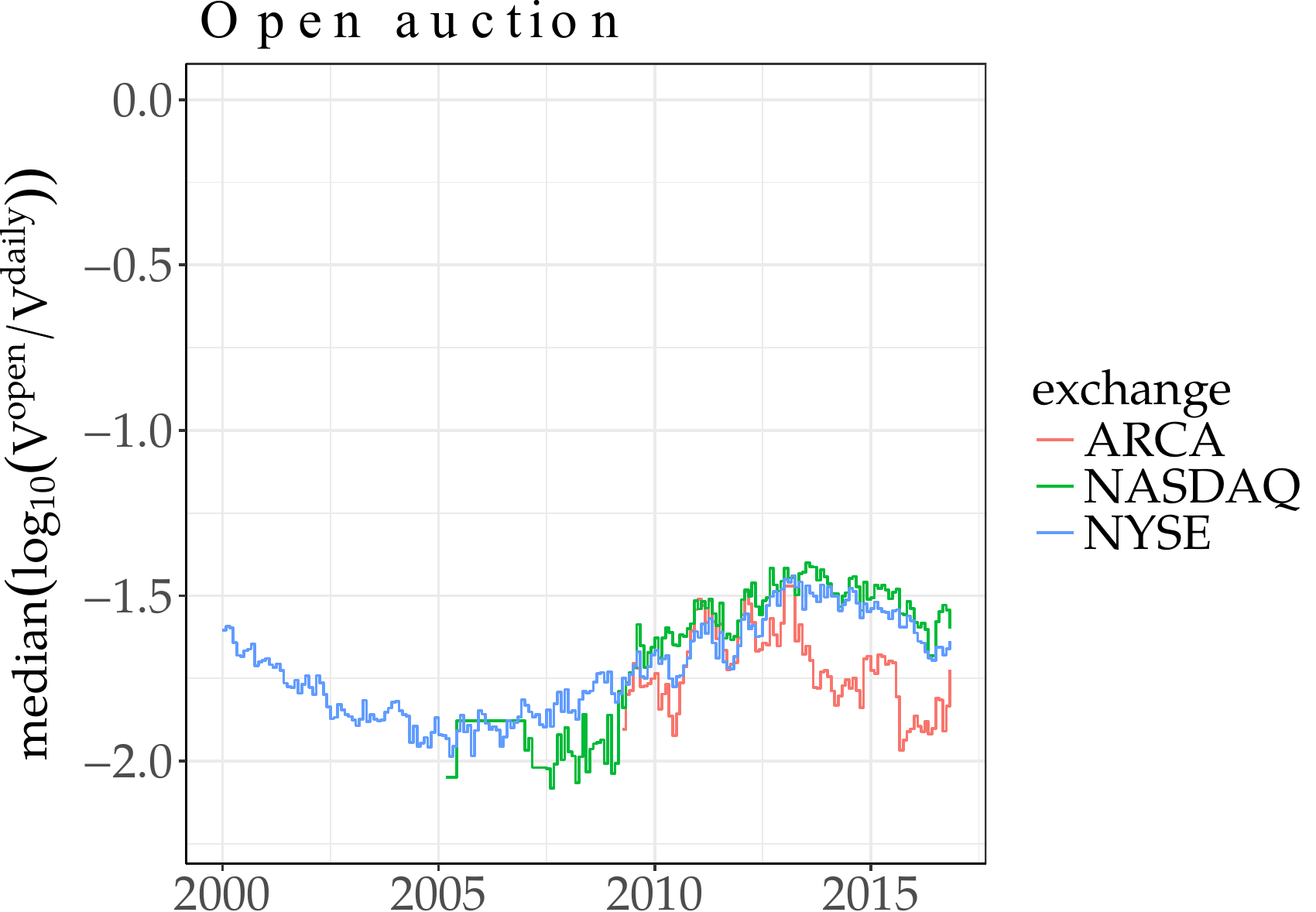}\includegraphics[width=0.5\textwidth]{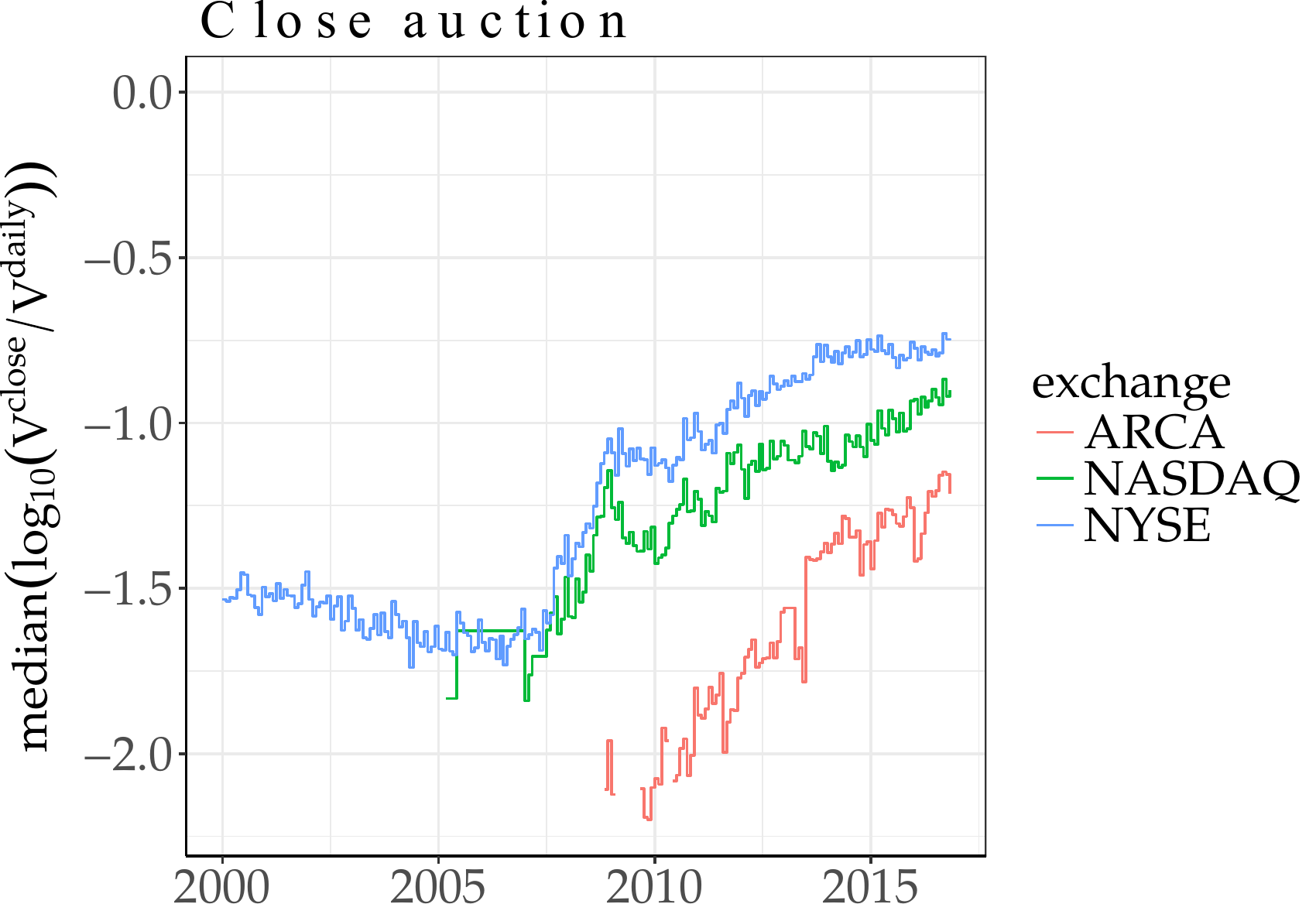}\caption{Monthly median of the volume exchanged at an auction divided by the
total daily volume; months with less than 100 assets have been filtered
out; left plot: open auction, right plot: close auction. \label{fig:frac_vs_time}}

\end{figure}
This, however, is a static picture determined over more than 7 years.
The fraction of volume exchanged at auctions is not roughly constant.
Figure \ref{fig:frac_vs_time} plots the monthly median fraction of
the auction volume divided by the total daily volume as a function
of time (months with less than 100 assets have been filtered out which
essentially removes the beginning of the time-series for ARCA and
NASDAQ. It turns out that the median closing auction fraction has
been steadily increasing for all exchanges, except for NYSE for which
this quantity has plateaued in 2015, while the opening auction fraction
has been decreasing since 2012, on average. One should note once again
that the median is taken over all assets of a given exchange, and
thus this plot does not apply to a given asset in particular, as the
difference between assets may be quite large. In other words, inference
should be performed asset by asset, which is beyond the scope of this
paper.

\section{Pre-auction dynamics}

Exchanges disseminate pre-auction information about indicative price,
current imbalance at that price and matched volume at that price.
The frequency of update is quite variable: ARCA gives away the most
frequent information, NYSE gives updates with increasing frequency
as the auction nears, while NASDAQ publishes  the smallest amount of information
(only one update before the open). We thus shall focus on
assets traded on ARCA and briefly on those traded on NYSE. We remind the notations: the imbalance
of asset $\alpha$ at auction $x$ on day $d$ at time $t$ is denoted
by $I_{\alpha,d}^{x}(t)$, the currently matched volume by $W_{\alpha,d}^{x}(t)$,
and the indicative price by $\pi_{\alpha,d}^{x}(t)$. 

\subsection{Typical activity}

\begin{figure}
\includegraphics[width=0.5\textwidth]{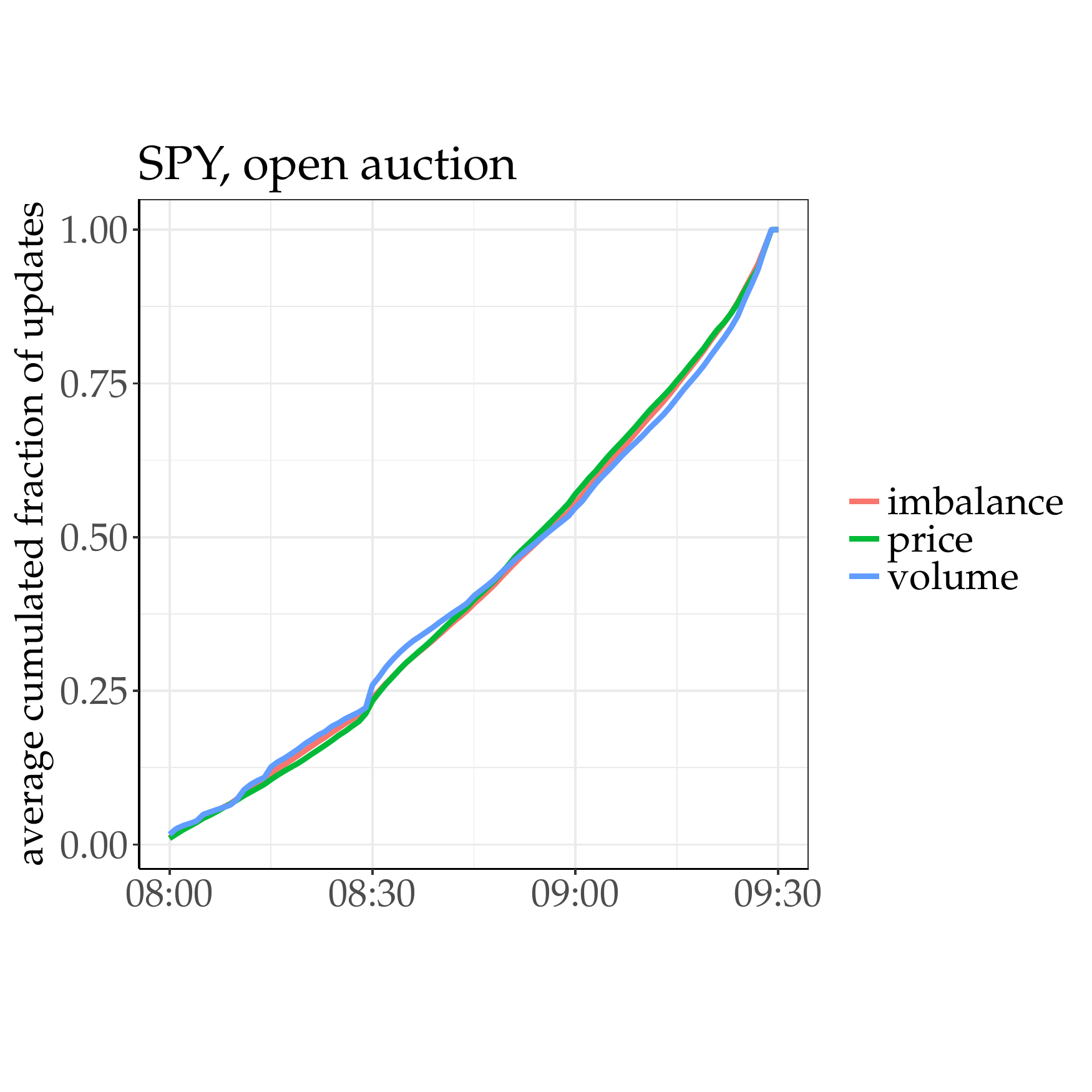}\includegraphics[width=0.5\textwidth]{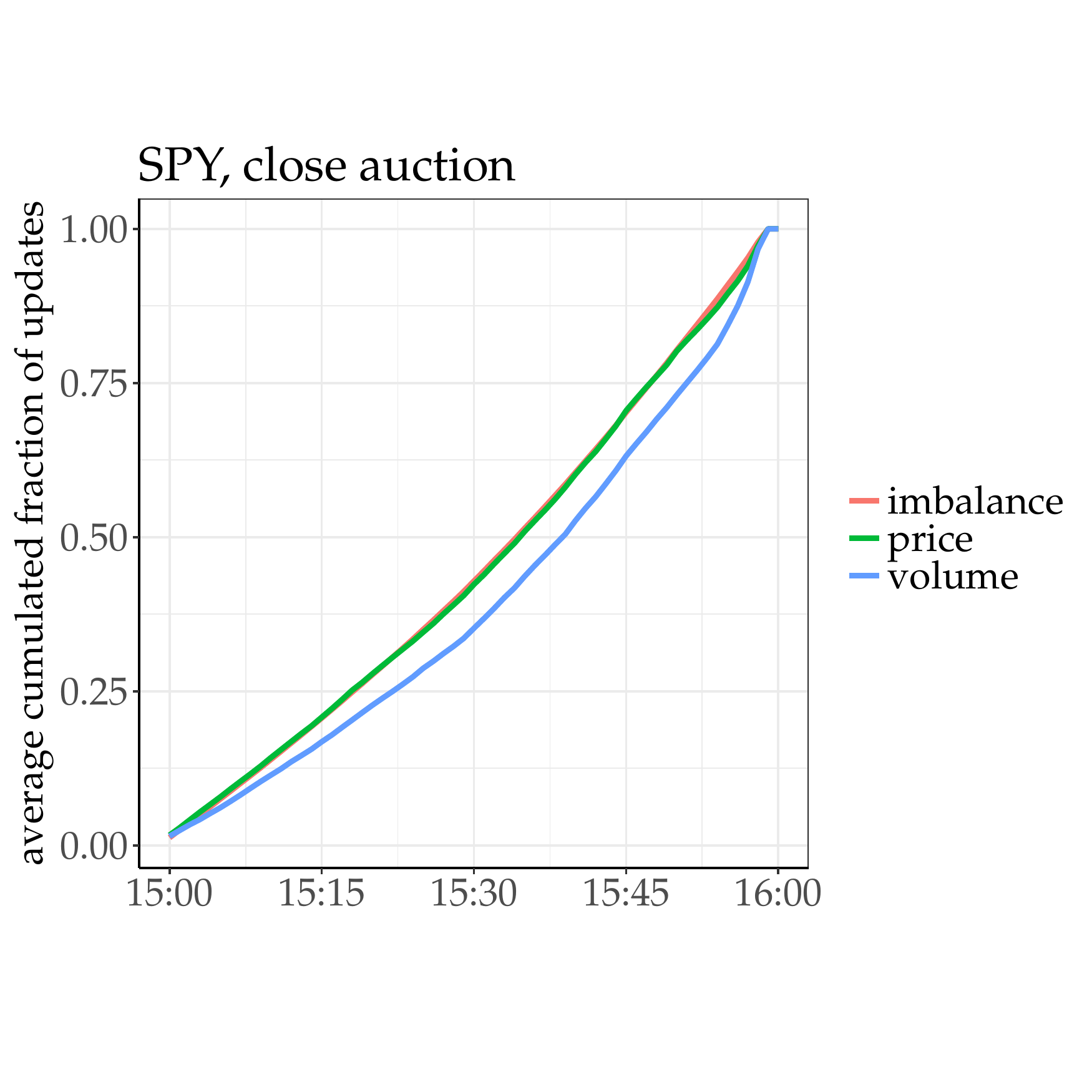}

\includegraphics[width=0.5\textwidth]{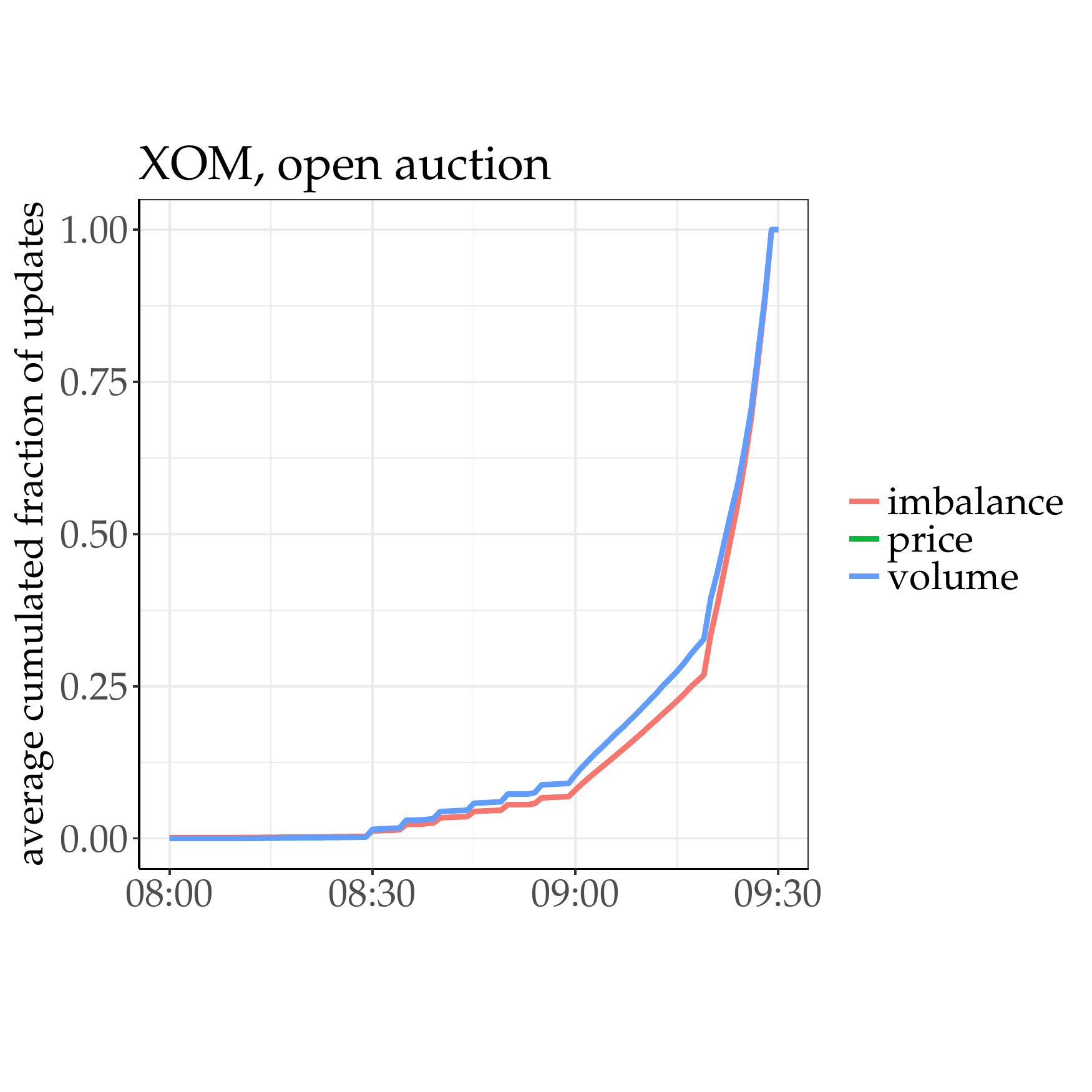}~\includegraphics[width=0.5\textwidth]{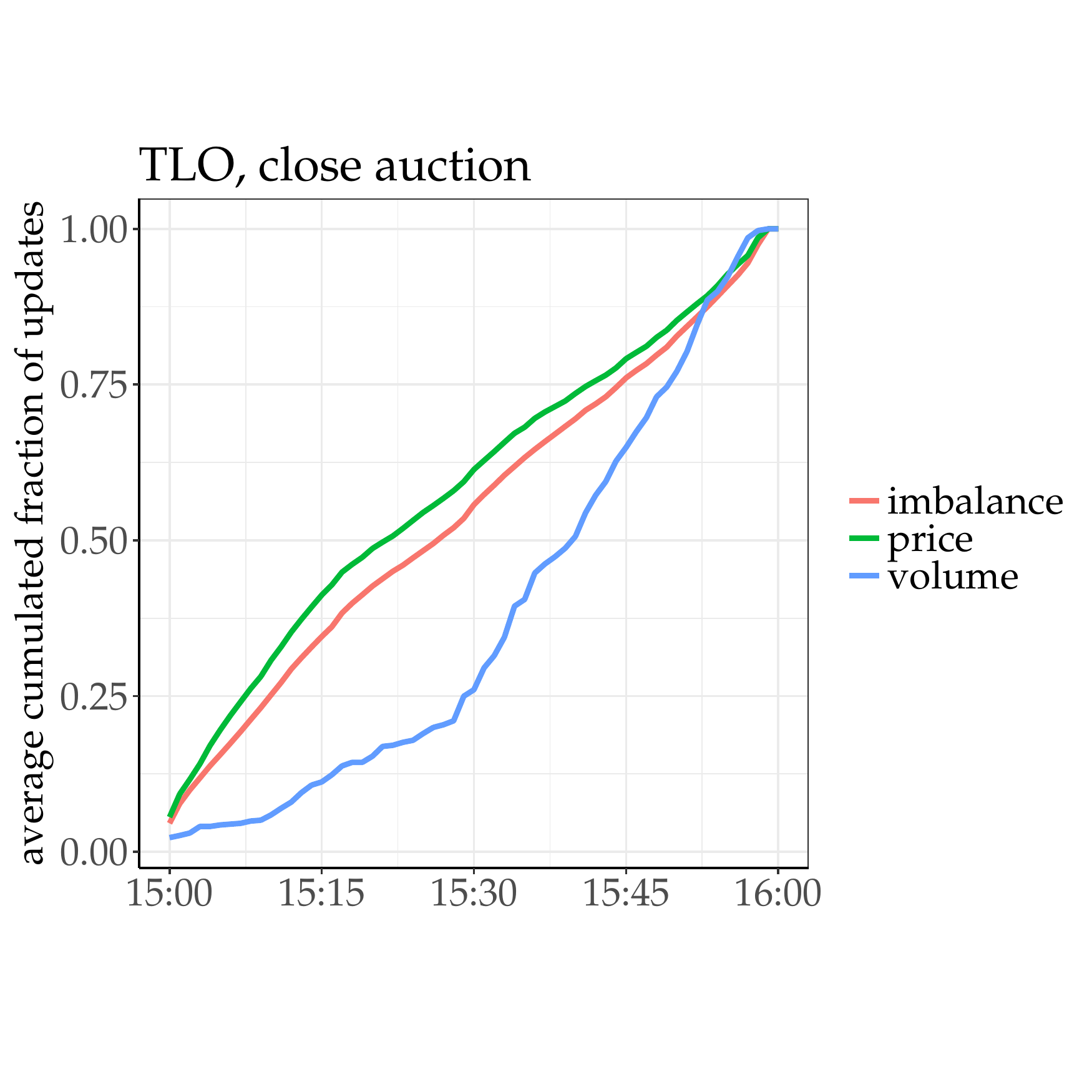}

\caption{Average fraction of cumulated number of events as a function of the
time for SPY (ARCA), XOM (NYSE) and TLO (ARCA). \label{fig:frac_updates}}
\end{figure}

Let us first investigate the typical update patterns. For each asset
and day, we measure at fixed time resolution (1 minute) the ratio
between the number of updates up to time $t$ (in minutes) and the
total number of updates for that particular day, and then compute
the average of this quantity over all the days. By definition, this
quantity increases monotonically. 

In the NYSE, updates to each quantity are disseminated at fixed times
with a strongly accelerating pattern (convex cumulated number of
updates), as can be seen for XOM in Fig.~\ref{fig:frac_updates}.
Update times of ARCA on the other hand are not fixed, which allows
one to quantify asset by asset how the activity unfolds, on average.
Intuitively, one would expect that the activity increases as the time
of the auction nears. This is the case in other kind of auctions with
fixed ending times, such as on eBay (see e.g. \cite{borle2006timing}).
Indeed, in auctions, sending a non-cancellable bid reveals some information about
one's intentions. However, the fact that activity may concentrate
just before the time of the auction is not always related to strategic
behavior. Indeed, human beings tend to prefer to act just before a
deadline in less competitive contexts, such as sending an abstract to a conference,
or paying its fee \citep{alfi2007conference,alfi2009people}. {\color{black}Delaying one's order submission minimizes the risk of divulging directional information too early. Another possibility to avoid leaking too much information for the traders  is to synchronize their actions at fixed times so that individual actions are lost in a multitude of orders, as it is the case for example at 8:30 for SPY (see below), and, to much larger extent in Paris Stock Exchange \citep{challet2018strategic}.
}

\begin{figure}
\includegraphics[width=0.5\textwidth]{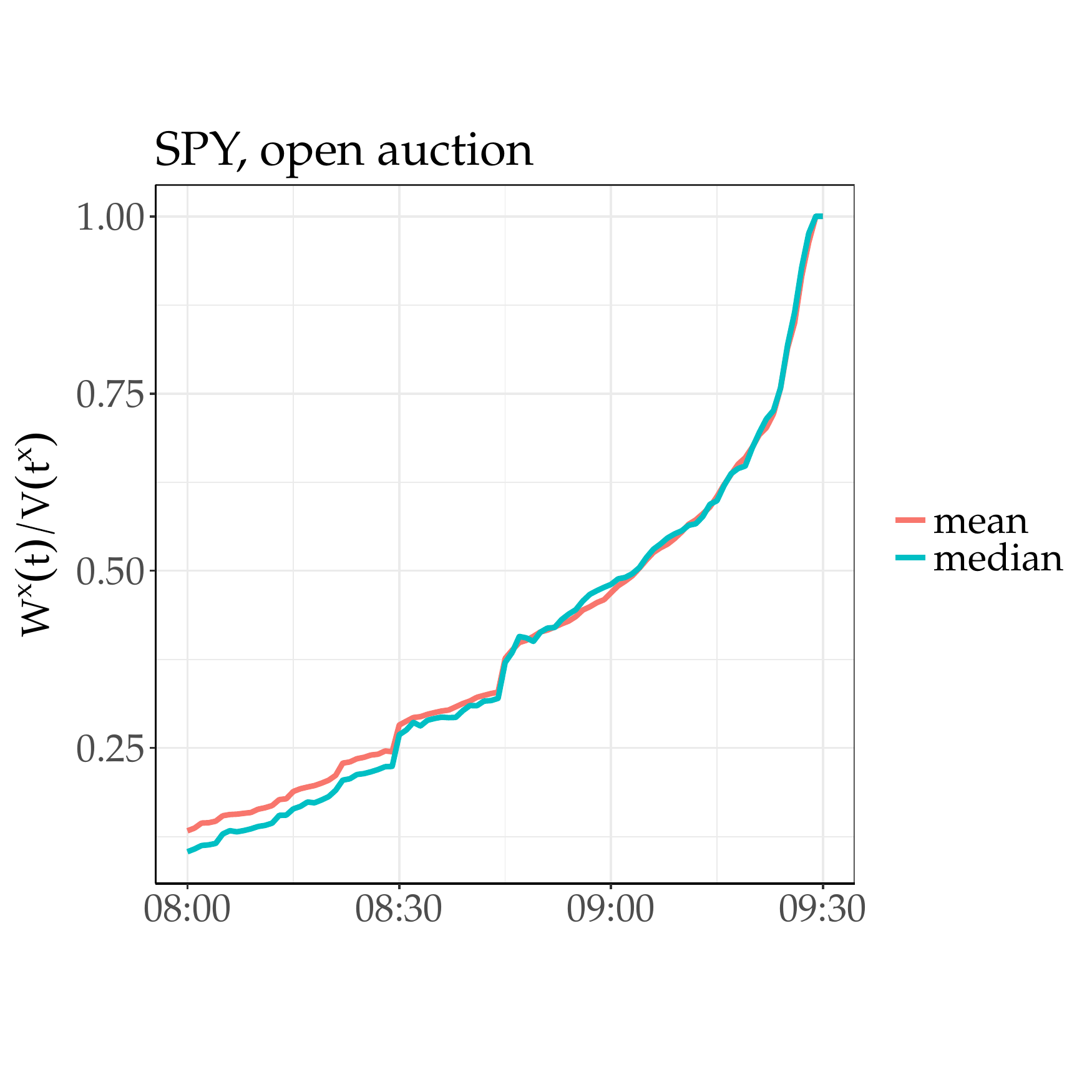}\includegraphics[width=0.5\textwidth]{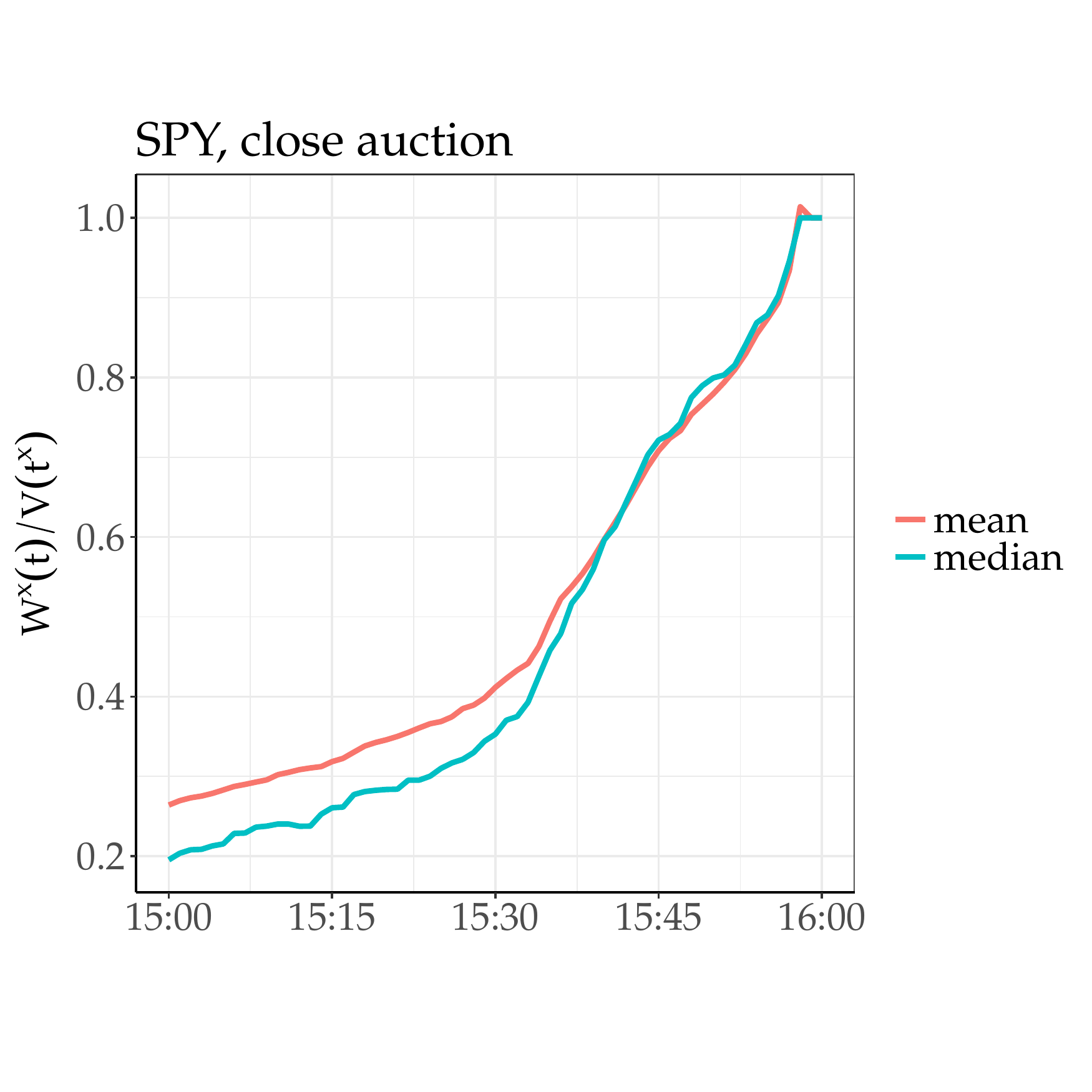}

\includegraphics[width=0.5\textwidth]{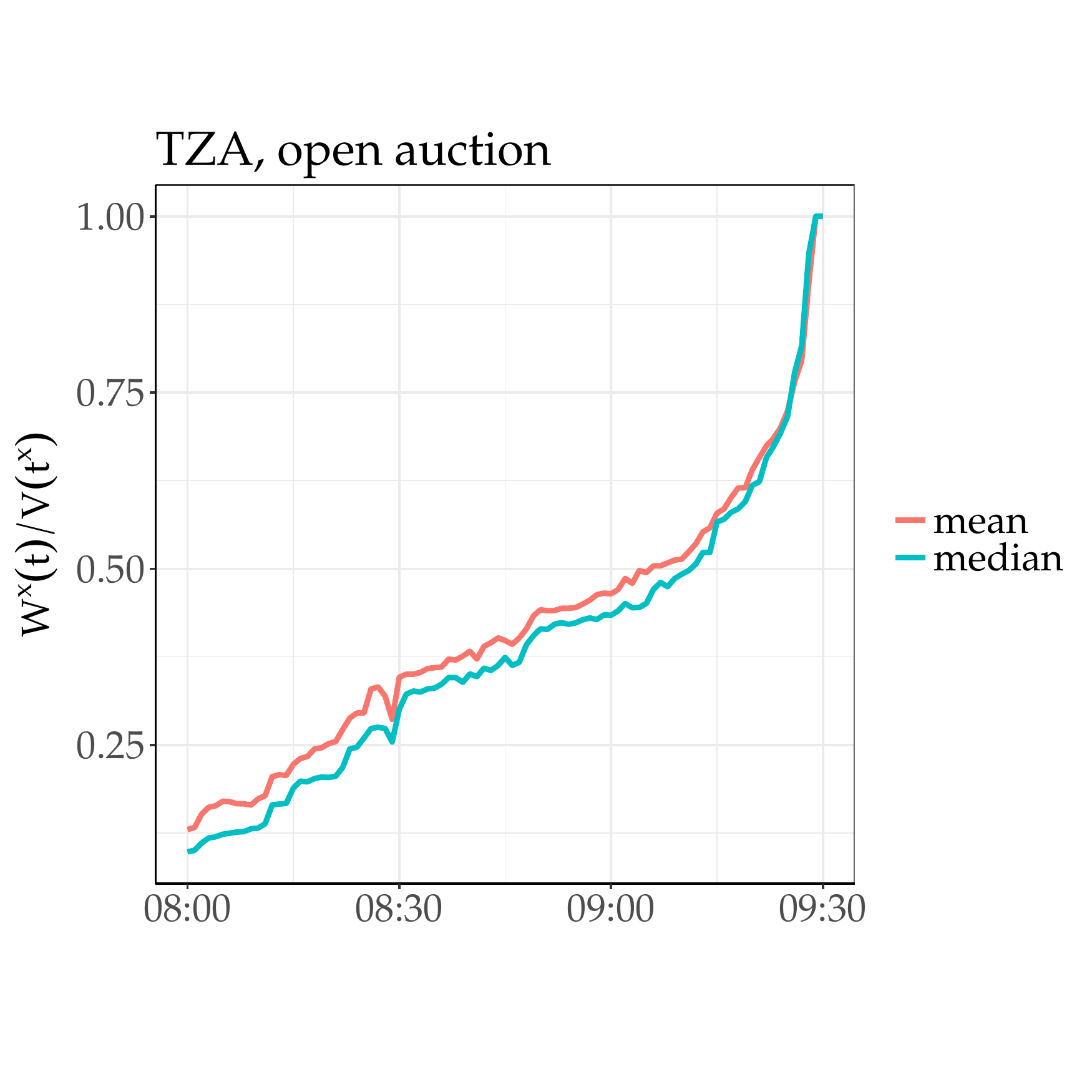}\includegraphics[width=0.5\textwidth]{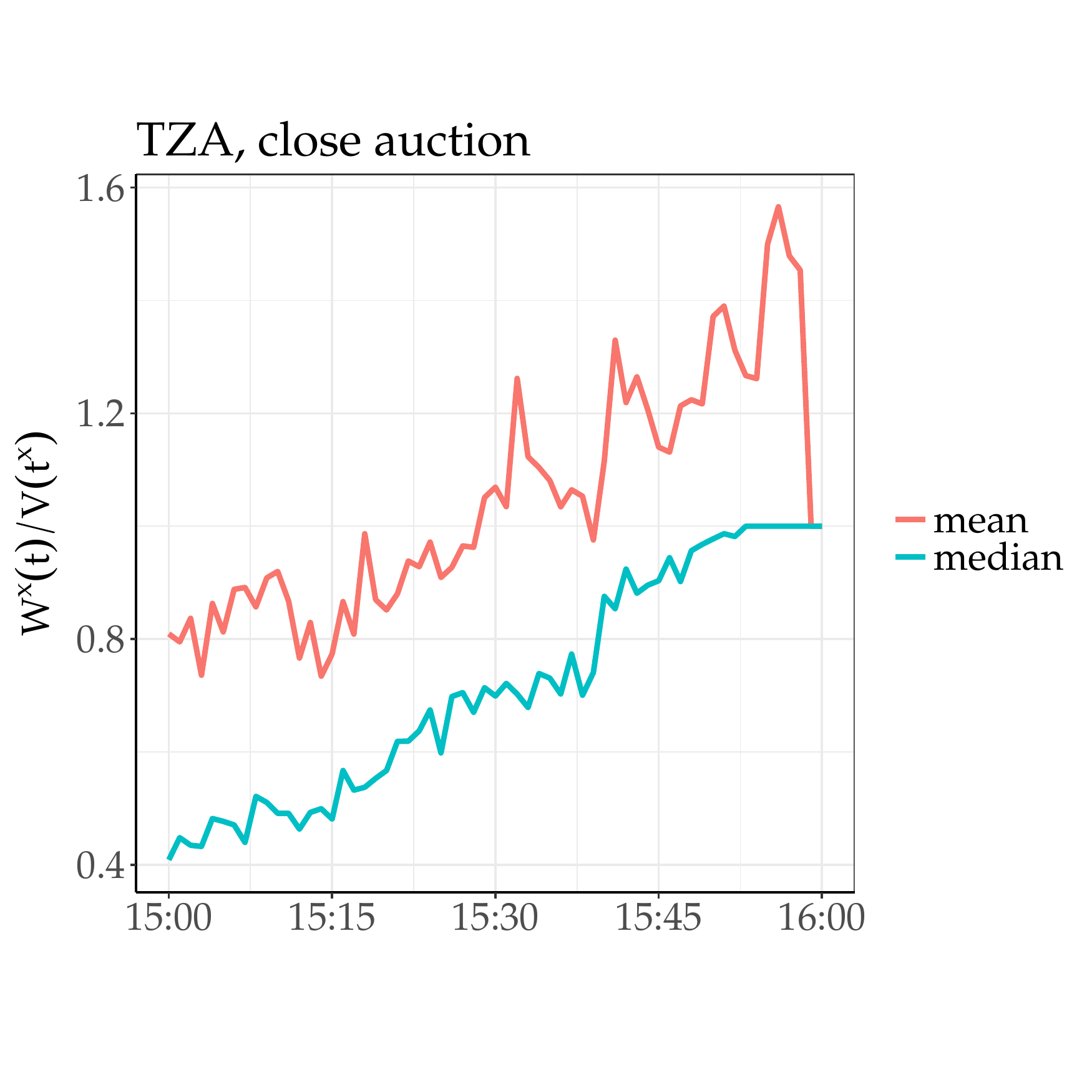}

\caption{Average and median fraction of matched volume and total auction volume
as a function of the time for SPY and TZA, traded on ARCA.\label{fig:fraction_total_volume_vs_t}}
\end{figure}

Most practitioners however would be more interested in the evolution
of the typical fraction of the indicative matched volume divided by
the final auction volume at a given time. Here, we compute for asset
and each day the fraction $W_{\alpha,d}^{x}(t)/V_{\alpha,d}^{x}$
every minute and compute its average and median for each asset. Figure
\ref{fig:fraction_total_volume_vs_t} plots this quantity for SPY
and TZA. During the opening auction, one generically notices an acceleration
of the matched volume which is much larger than that of the number
of volume updates, especially after 9:15. This implies that larger
orders are submitted closer to the auction ending time, which is consistent
with people trying to avoid having too much direct impact on the indicative
price since the matched volume increases as a function of time. One
also notices that a sizeable fraction of the daily volume auctions
are submitted at or near 8:30 and 8:45 for both assets. Nevertheless,
these two tickers have been chosen to illustrate the lack of universality
across assets. Indeed, TZA clearly shows that during some days, a
large fraction of the auction is cancelled just before the closing
auction, but that this is not systematic as the median fraction does
not exceed 1. This also happens for other assets either at the opening
or closing auctions. According \cite{bellia2016low},
cancellations near auction ending time are mostly due to high-frequency traders. {\color{black} The behavior of this quantity on US exchanges is markedly different from that of Paris Stock Exchange for which \cite{challet2018strategic} finds that the number of events increases as a power-law until the auction time.}

While SPY has a clearly accelerating, convex average matched volume
ratio as a function time, we systematically investigated this pattern
asset by asset by fitting polynomials of the first and second 
degree to $W_{\alpha,d}^{x}(t)/V_{\alpha}^{x}(d)$, and computing
their respective Akaike Information Criterion (AIC). Using once again
the Vuong ratio test, one finds that a linear fit is favored in 3\%
for the assets at the opening auctions, and none at the closing auction
at the 5\% level; reciprocally, a second degree fit is better than
a linear fit in 64\% for both auctions, the remaining 33\% being undecidable.
Among these 64\% of assets, 75\% have a convex, accelerating behavior
at opening auctions, and 89\% at closing auctions.

Finally, one may be interested in the typical time at which the average indicative
matched volume reaches  a given percentage of the auction volume.
For each asset traded on ARCA, we computed the median time (over the
days) of the time at which 50\% of the total auction volume was matched and
found two peaks centered at 8:30 and 8:55 for the opening auction,
and a large peak centered at 15:35 for the closing auctions.

\subsection{Mean-reversion and sub-diffusive prices}

\begin{figure}

\includegraphics[width=0.35\textwidth]{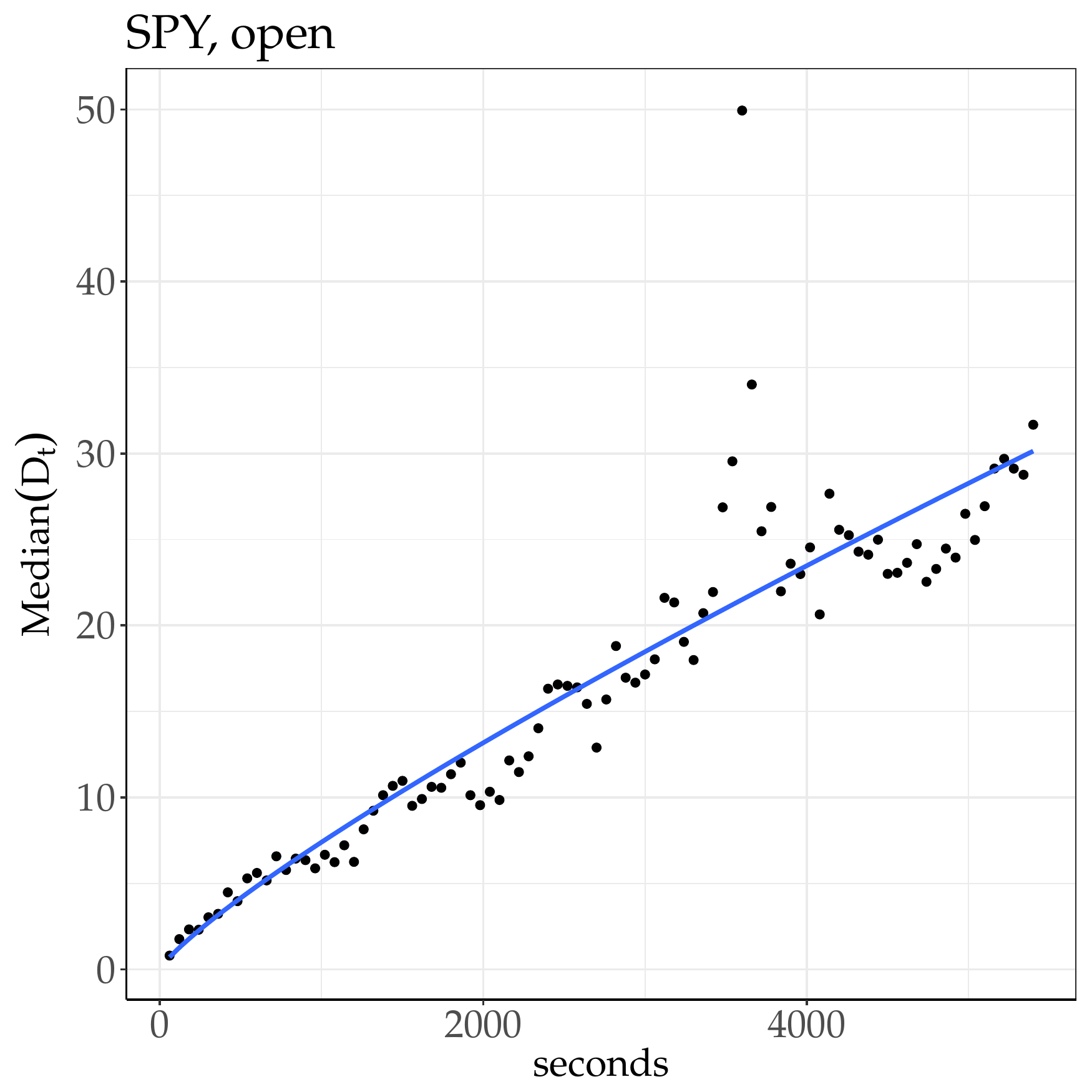}\includegraphics[width=0.35\textwidth]{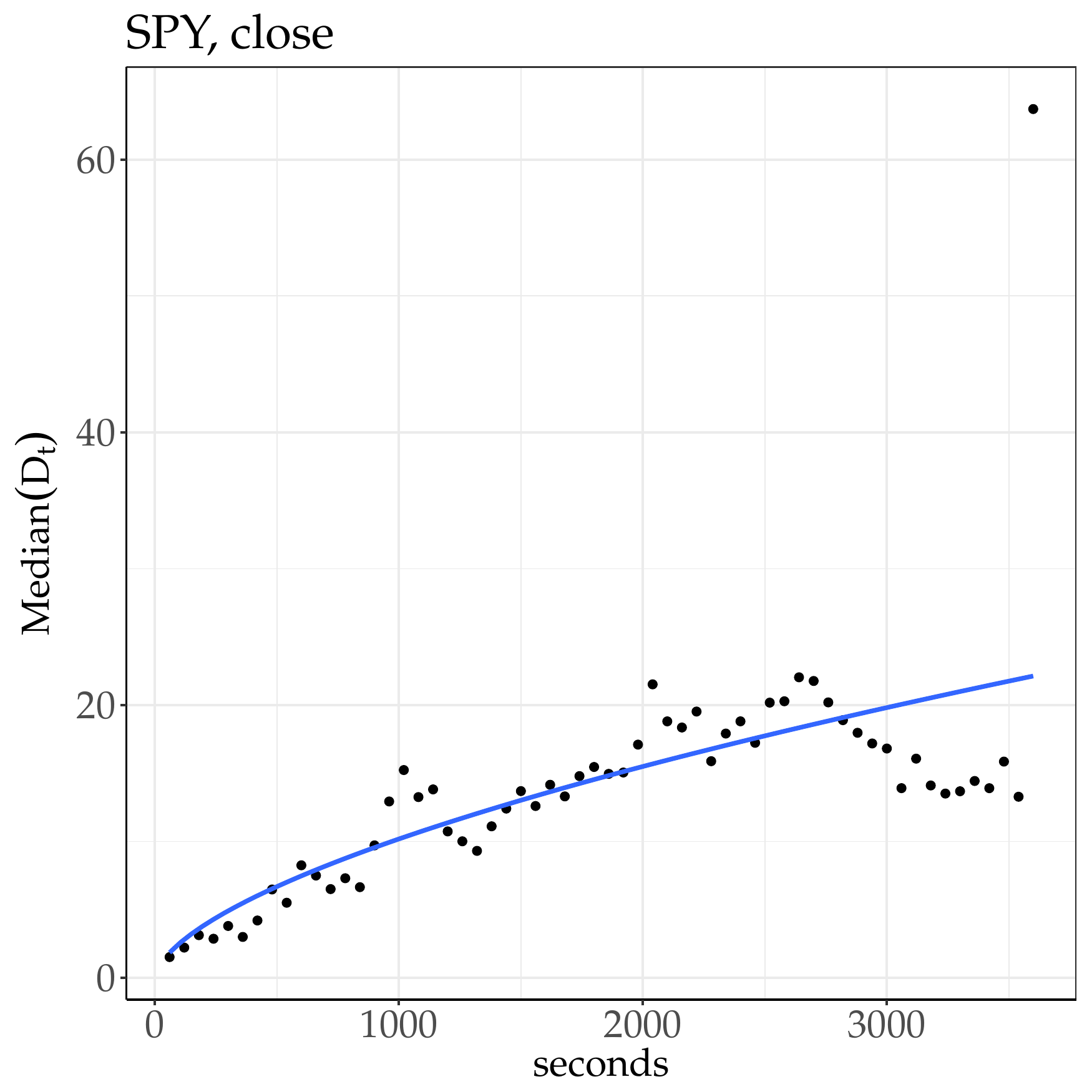}\includegraphics[width=0.35\textwidth]{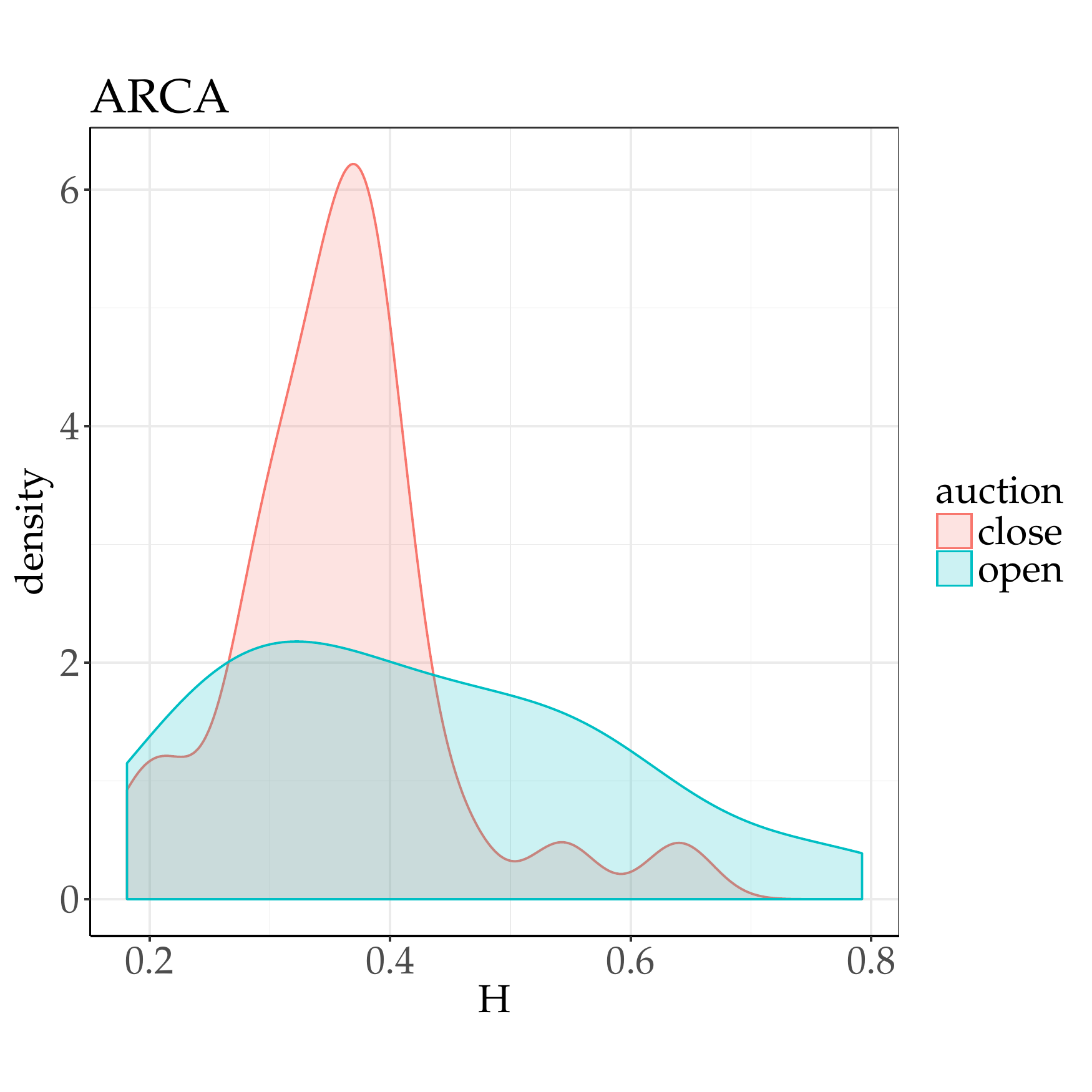}\caption{Left and center plots: median square dispersion of the indicative
auction price with respect to the auction price versus the time remaining
until the auction, together with a power-law fit (Hurst exponent $H\simeq0.45$
for the opening auction and $H\simeq0.23$ for the closing auction.
Right plot: histogram of the Hurst exponent $H$ for all tickers of
ARCA from fits whose p-values associated with $H$ are smaller than
0.001. \label{fig:Hurst_price}}
\end{figure}

Focusing on ARCA data, we first check if the indicative price behaves
as a standard Brownian motion, or if the existence of a final auction
time and the fact that the activity and matched volume increases steadily have
an influence of the indicative price fluctuation patterns. More precisely,
we measure how the squared difference between the indicative match
log-price $\log \pi_{\alpha,d}^{x}(t)$ and the final auction log-price
$\log p_{\alpha,d}^{x}(t^{x})$, rescaled by daily variance of the indicative
price log returns, scales with $\tau=t^{x}-t$ by defining 
\begin{equation}
D(\tau)=\frac{\log(p_{\alpha d}^{x}(t^{x})/\pi_{\alpha,d}^{x}(t)){}^{2}}{\text{var}([\log(\pi_{\alpha,d}^{x}(t)/\pi_{\alpha,d}^{x}(t-1))])}\label{eq:Hurst}
\end{equation}

Generally, $D(\tau)\propto\tau^{2H}$ where $H$ is the Hurst exponent.
Diffusive processes have $H=1/2$, whereas under-diffusive processes
correspond to $H<1/2$. Note that we normalize by the variance of
the indicative price log returns of each day, so as to be able to
define $D$ as an average over days in a meaningful way. We first filter out single auctions with less than 50 price updates. We then
discretize $\tau$ into slices of 1 minute, compute the median of
$D$ of the days for each slice $\tau$, and perform a non-linear
fit $\text{median }D(\tau)=a\tau^{2H}$ (removing $\max_{\tau}\text{median }D(\tau)$
for the sake of robustness). The left and middle plots of Fig.~\ref{fig:Hurst_price}
show an example of the dependency of $D$ on $\tau$ for the asset
with the largest number of updates (SPY); the peak at 08:30 is due
to the fact that many traders submit auction orders around that
time, whose effective imbalance may considerably shift the indicative
match price, usually followed by an opposite jump or gradual relaxation.
The right-hand-side plot displays the density of $H$ over all the
assets with enough data on ARCA (we only kept fits whose p-values
associated with $H$ are smaller than 0.001). We find that the indicative
price of the opening auction is sub-diffusive for 73\% of the assets
for the opening auction and 93\% of the assets for the closing auction. 

The sub-diffusive property probably has several causes. First, diffusion
is computed backwards from the auction ending time, thus the typical event rate decreases as the time-to-auction increases, which mechanically
leads to sub-diffusion, provided that the immediate impact per event
stays roughly constant. {\color{black}\cite{challet2018strategic} discusses this hypothesis in greater details with data from Paris Stock Exchange, in which the analysis is simplified by simple scaling laws of both the number of events and the typical indicative price change as a function of time, and concludes that mechanistic effects alone cannot account for the observed price behavior, which confirms the importance of strategic behavior. Since no such scaling law holds in our data, we cannot replicate this result for US equities.}

\begin{figure}
\includegraphics[width=0.35\textwidth]{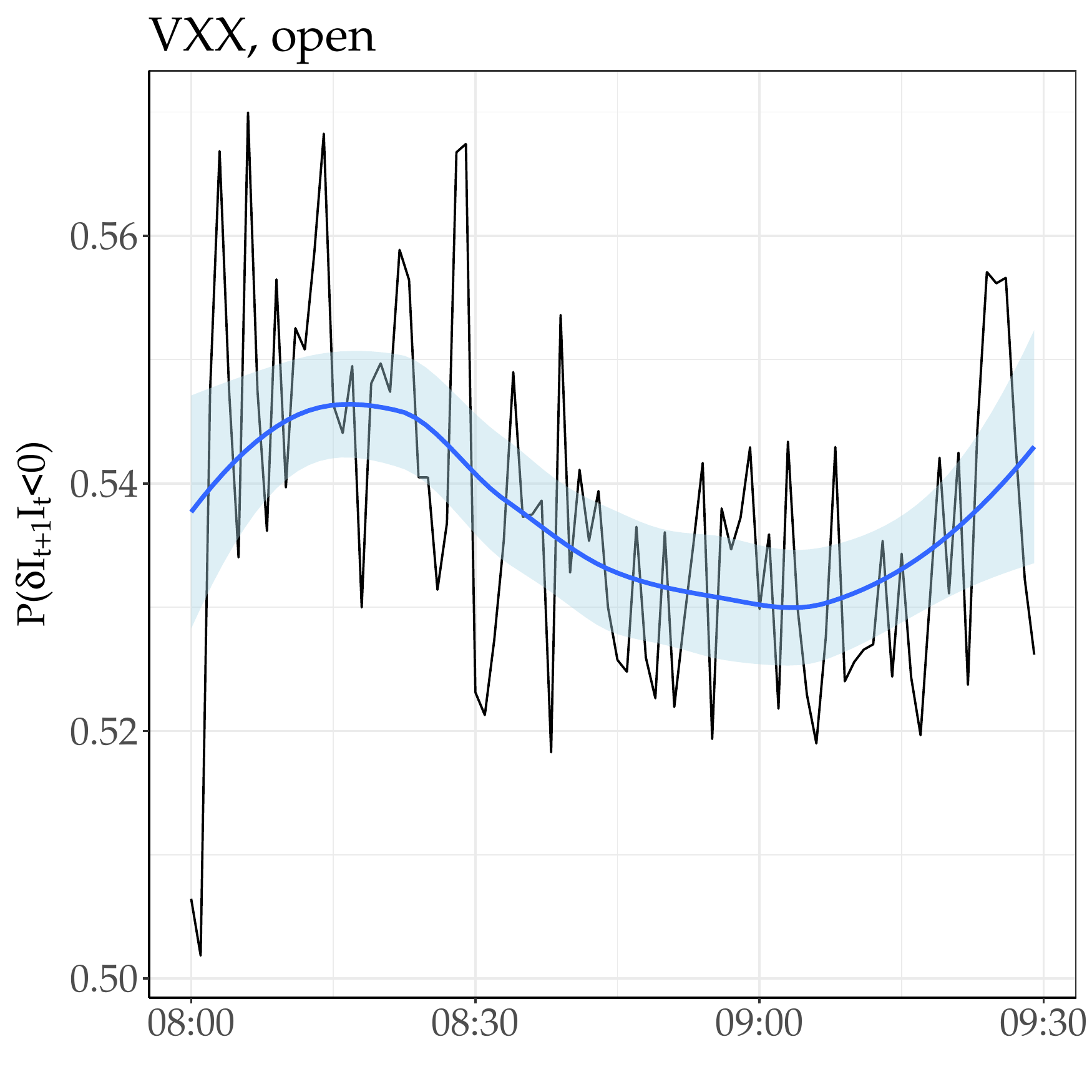}\includegraphics[width=0.35\textwidth]{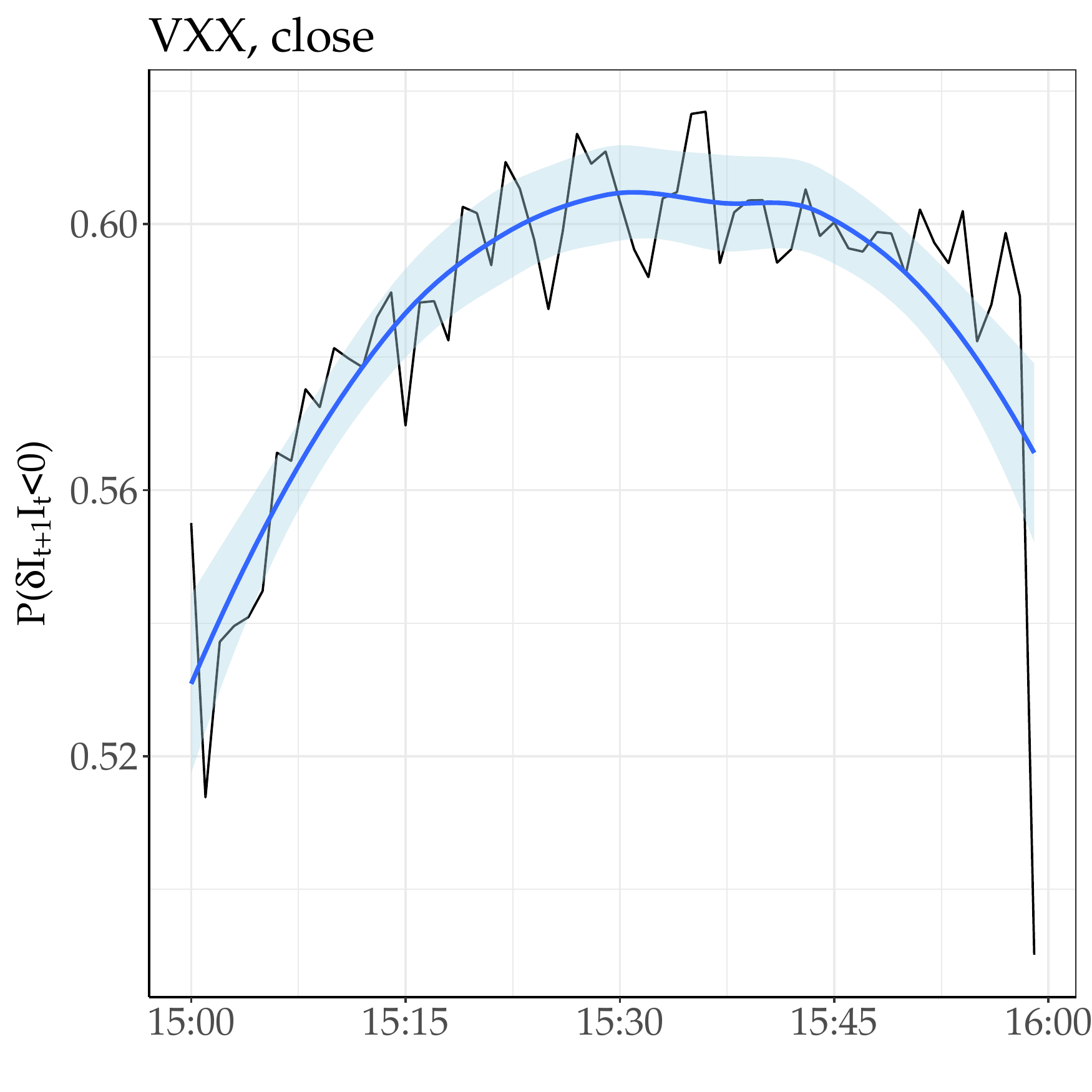}\includegraphics[width=0.35\textwidth]{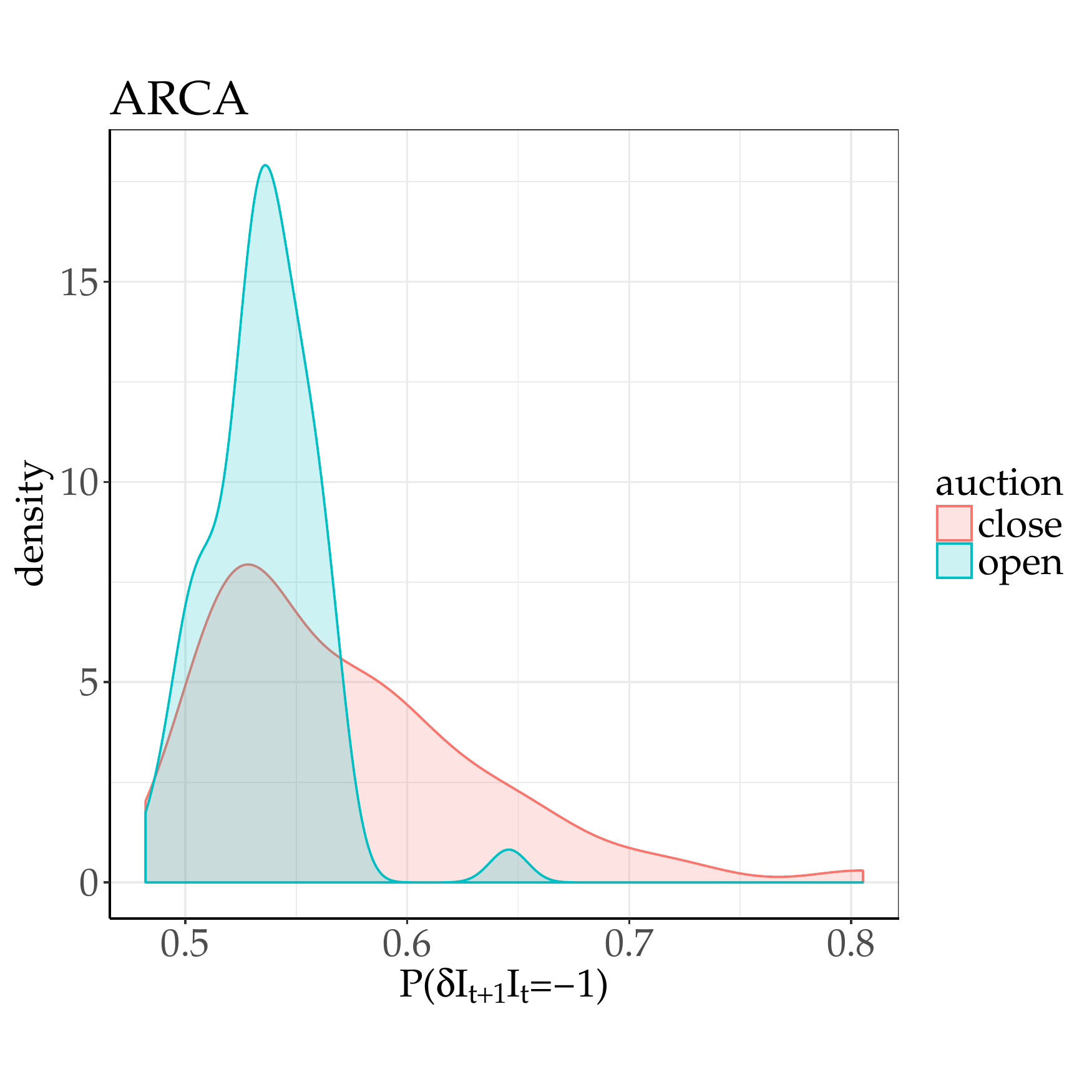}\caption{Left and middle plot: average probability that a new event reduces
the imbalance as a function of time for VXX and both opening and closing
auction. Right plot: histogram of the average over days for each ticker
of that probability.\label{fig:PsdII}}
\end{figure}

Indeed, another cause of sub-diffusion comes from the way new orders tend
to cancel the current imbalance on average. In other words, a strategic
way of placing new auction orders consists in waiting until the imbalance
sign is the opposite from that of one's intentions and only then sending
one's orders. Defining the imbalance difference of a new event (new
order or cancellation) by $\delta I_{t+1}=I_{t+1}-I_{t}$, this strategic
behavior can then be characterized by $P[\text{sign }(I_{t+1}\times\delta I_{t})=-1]$.
Note that this also includes order cancellation, which is much rarer
than order placement and will be thus neglected here. Figure \ref{fig:PsdII}
shows the evolution of this conditional probability as a function
of time for VXX, which is representative of many other assets whose
main exchange is ARCA: this probability is smaller for the opening
auction than the closing auction for about 95\% of the assets. In
addition this probability shows a clear maximum for the closing auction
at around 15:30. Some other assets display an increase of this probability
in the last few minutes before the auctions (SPY, for one). In passing,
note that this probability is about 1/2 for SPY during the opening
auction, and about 0.53 (and roughly constant) during the closing
auction, which is probably due to the fact that our data feed cannot
keep up with the event rates of this asset.

\subsection{Response functions}

A more refined way to characterize the dynamics of pre-auction periods
consists in measuring how the final auction price reacts to a given type of event as a function
of the time remaining until the auction. For example, is the average
response of the auction price to a new buy order placed at time $t$
positive or negative? In open-market order books, the answer is intuitive
enough. Auctions are different at least for two reasons. First, as
there is no trade before the auction ending time, real impact is not immediate.
Second, sending an order to an auction reveals some information about
one's intentions, thus there is a clear strategic aspect to the time
of order submission or cancellation. In other words, early orders
may trigger a different response than later ones simply because the
traders who submit the former have different strategies or expectations
than the latter ones and may or may not cancel some of their orders before the auction ending time, in line with \cite{boussetta2016role}.

Let us first derive the nature of several kinds of events from the
sign of the changes of indicative imbalance and matched volume. Table~\ref{tab:Type-of-event}
lists the three kind of event that may be inferred from out data feed.
Note that this kind of determination becomes imprecise if the fraction
of missing updates becomes substantial. 

\begin{table}[H]
\centering{}%
\begin{tabular}{|c|c|c|}
\hline 
$\delta I$ & $\delta M$ & event\tabularnewline
\hline 
\hline 
$>0$ & $>0$ & new buy order\tabularnewline
\hline 
$<0$ & $>0$ & new sell order\tabularnewline
\hline 
any & $<0$ & cancellation or limit order price revision  \tabularnewline
\hline 
\end{tabular}\caption{Type of event depending on the change of imbalance, matched volume
and match price. \label{tab:Type-of-event}}
\end{table}

\begin{figure}
\includegraphics[width=0.5\textwidth]{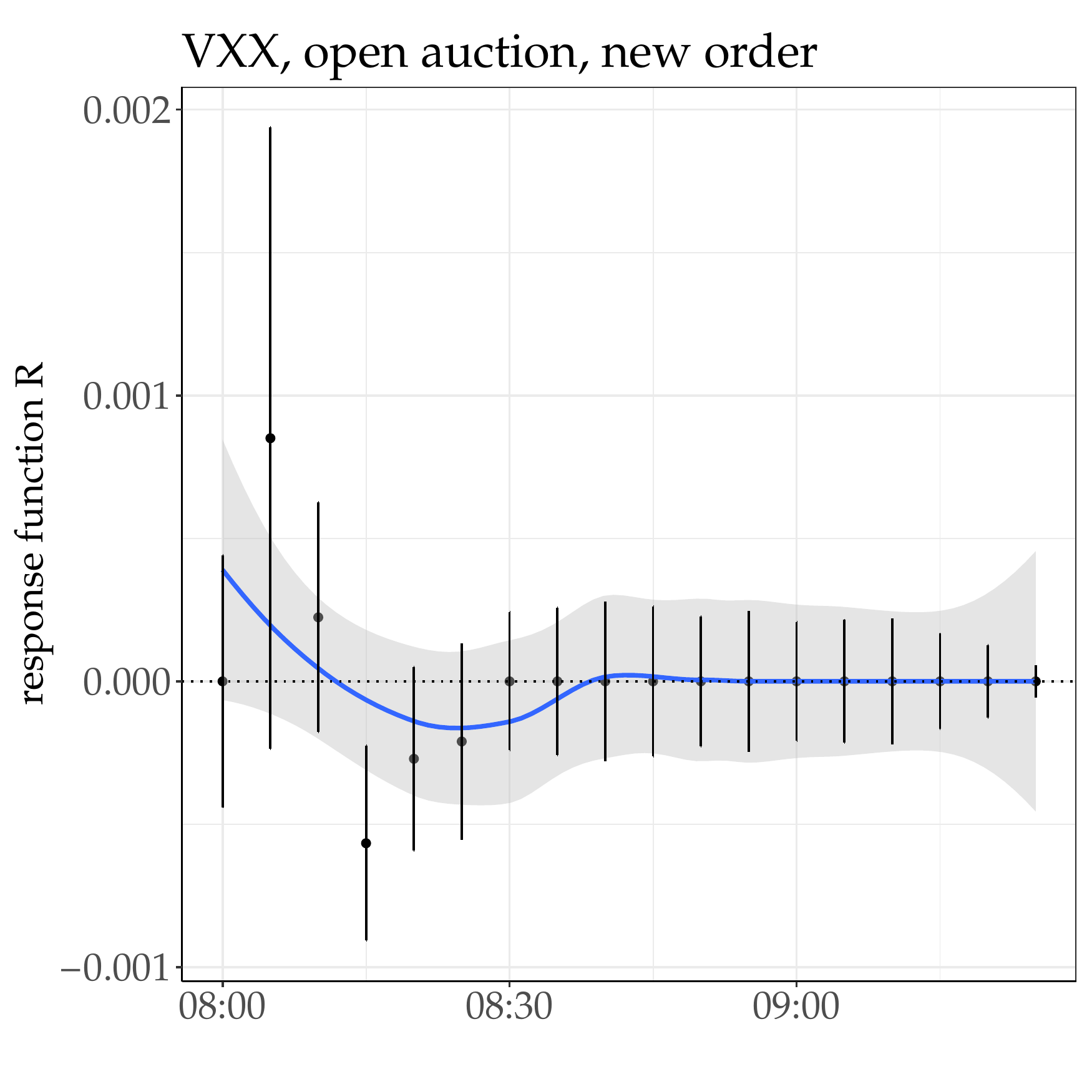}\includegraphics[width=0.5\textwidth]{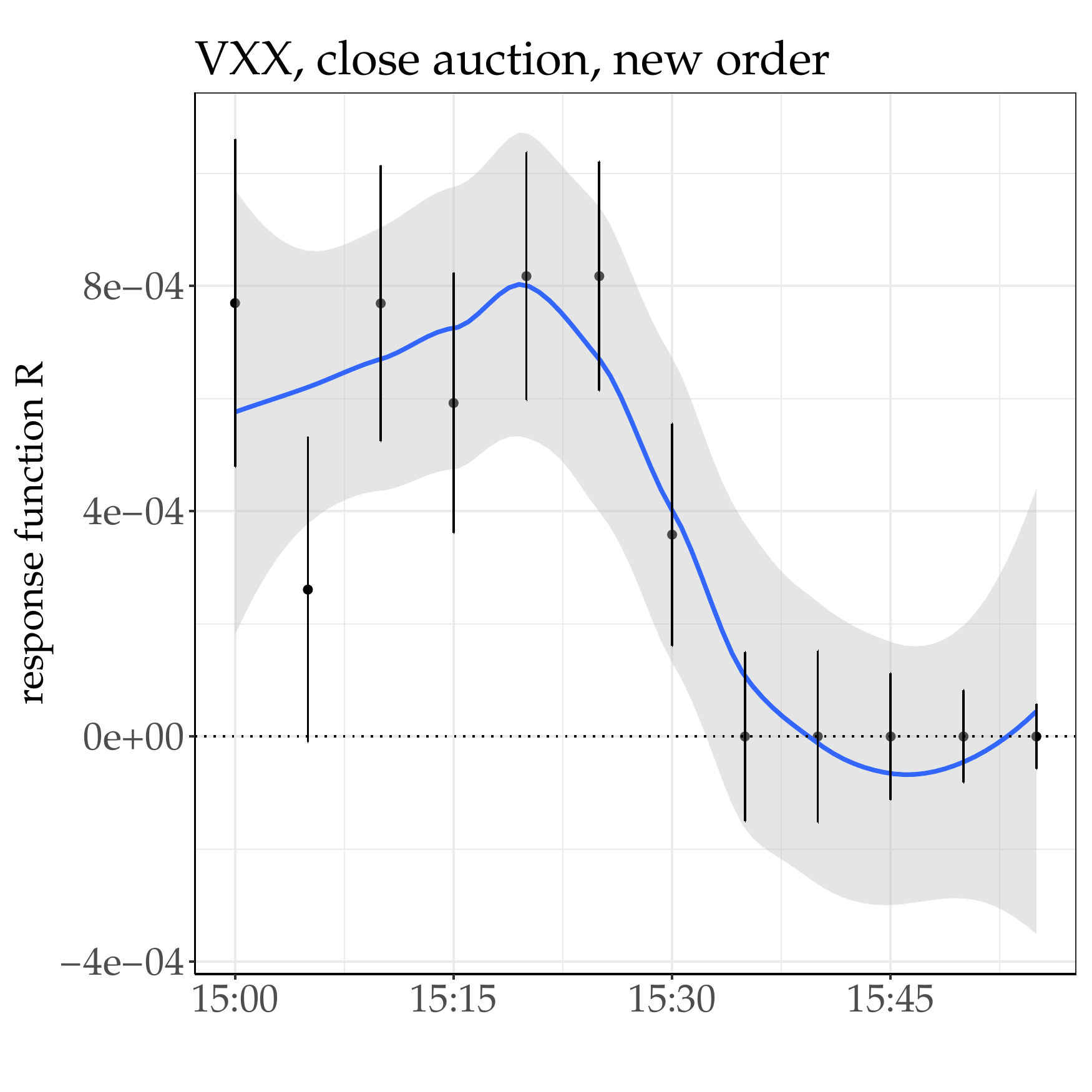}

\includegraphics[width=0.5\textwidth]{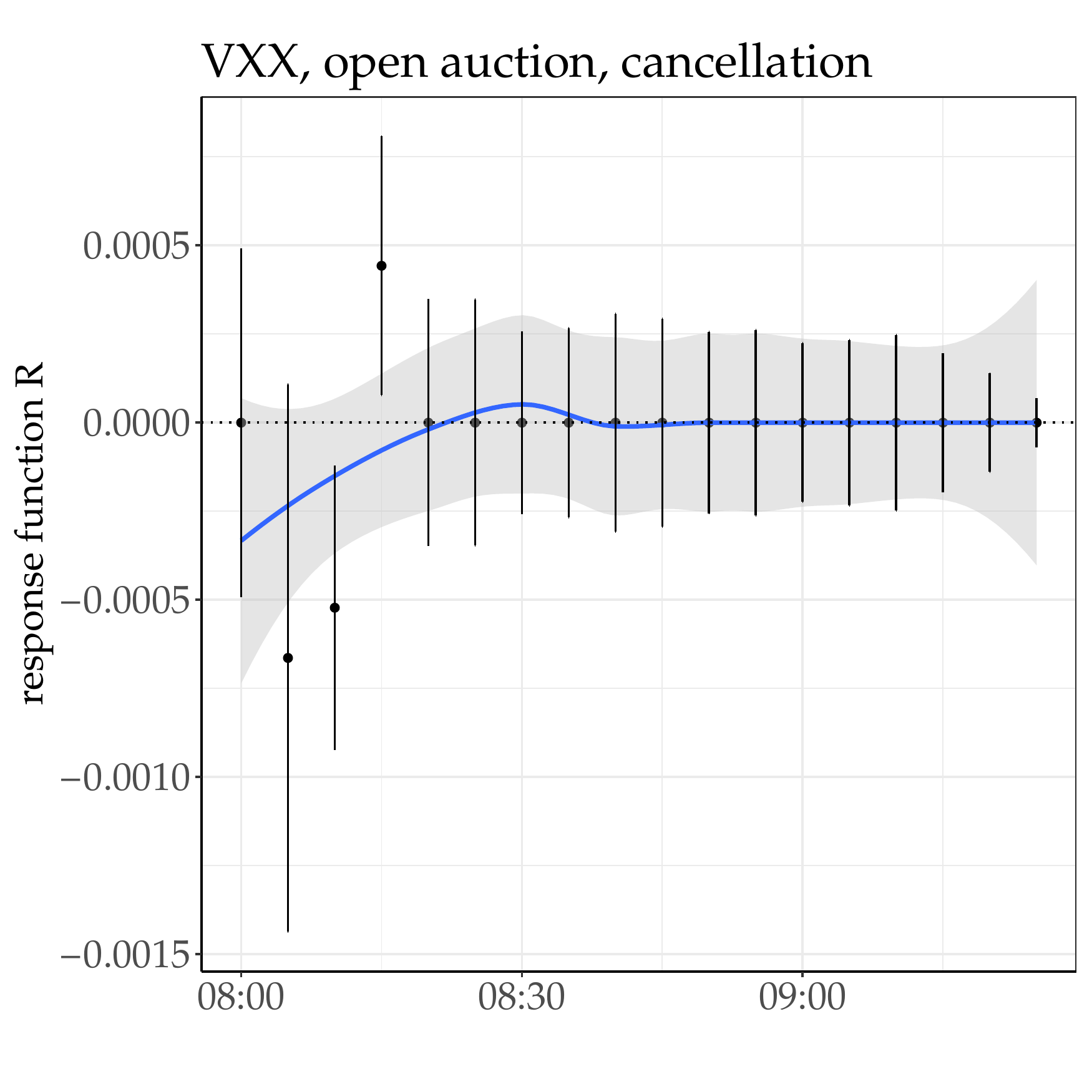}\includegraphics[width=0.5\textwidth]{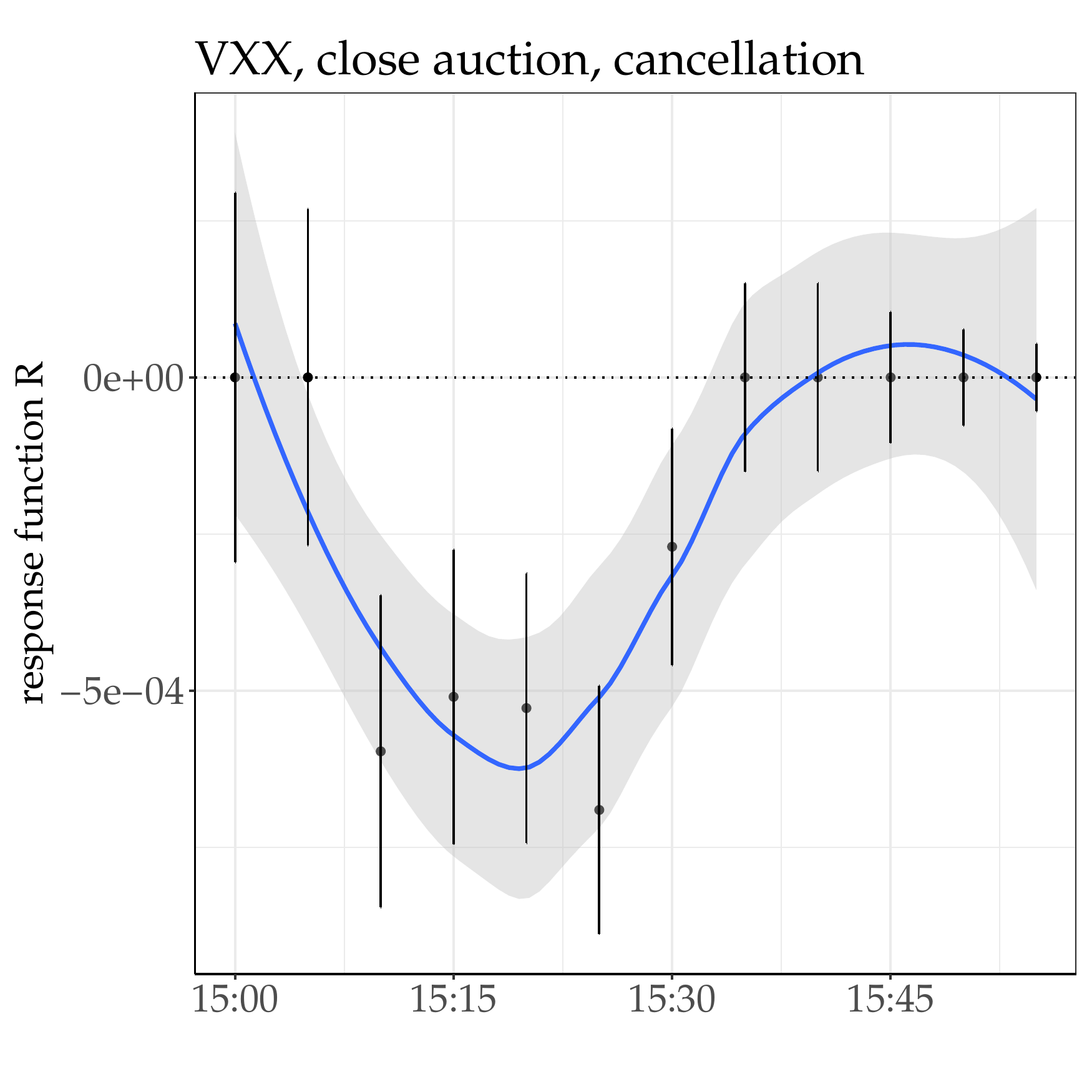}\caption{Median response functions of a new order (left plot) and order cancellation
during the auctions of VXX.\label{fig:Response-functions} Error bars
correspond to two standard deviations.}
\end{figure}

Inspired by \cite{bouchaudlimit4}, we define the linear response
function of the auction price to a new order, for a given asset $\alpha$,
at given update time $t_i$, as
\begin{equation}
R_{\text{\ensuremath{\alpha},+}}(t_{i})=\text{median}\left(\epsilon_{\alpha,d}(t_{i})\left[\log p_{\alpha,d}^{x}-\log\pi_{\alpha,d}^{x}(t_{i})\right]|\delta W_{\alpha,d}(t_{i+1})>0,\delta I(t_{i+1})\ne0\right)_{i,d},\label{eq:R}
\end{equation}

where $\epsilon_{\alpha,d}(t_{i})$ is the sign (1 for buy and -1
for sell orders) of the $i-$th order placed during day $d$ for auction
$x$ at time $t_{i}=t$ and asset $\alpha$, and $p_{\alpha,d}^{x}$
is the auction of price at time $t_{x}$. The idea is that on average
a new buy order pushes the auction price $p^{x}$ in direction opposite
to that of a sell order, hence the multiplication of the price difference
by the sign of the new order. The same kind of response function can
be defined for order cancellations, which occur when the matched volume
decreases ($\delta W<0$), which we denote as $R_{\text{\ensuremath{\alpha},-}}$.
Figure \ref{fig:Response-functions} shows that early new orders have
a definitely positive impact on average, while cancellation has an
opposite effect. 

\begin{figure}
\includegraphics[width=0.5\textwidth]{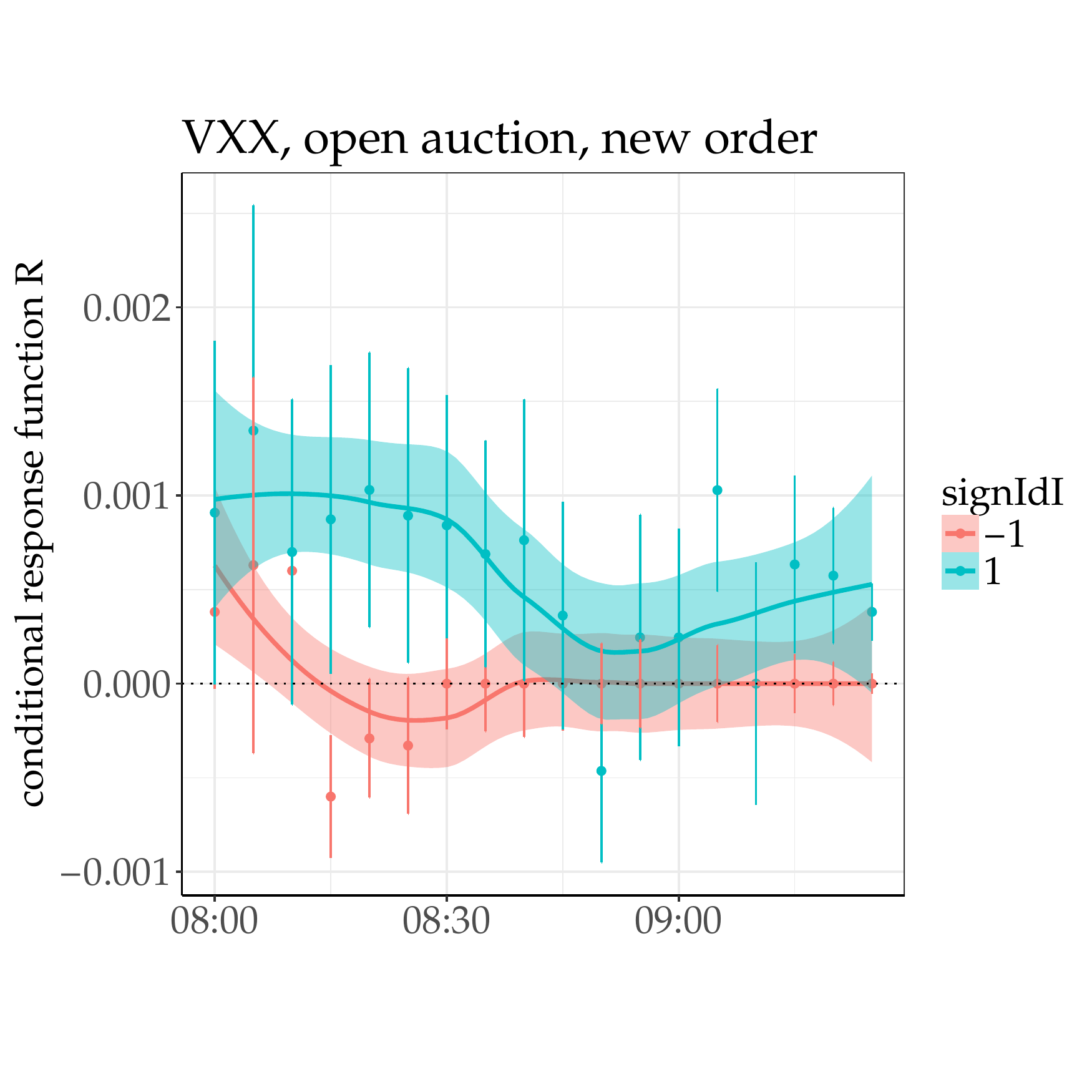}\includegraphics[width=0.5\textwidth]{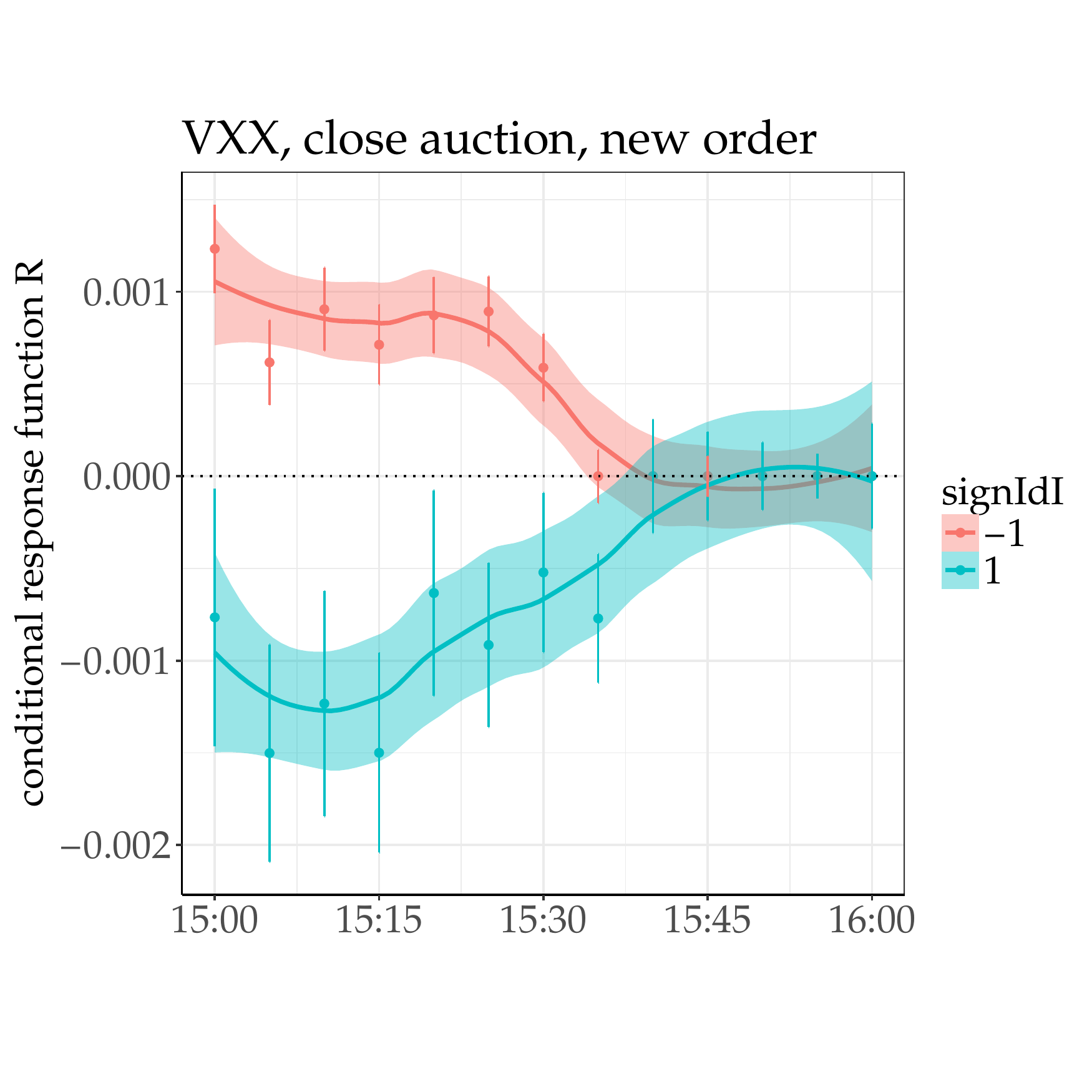}

\includegraphics[width=0.5\textwidth]{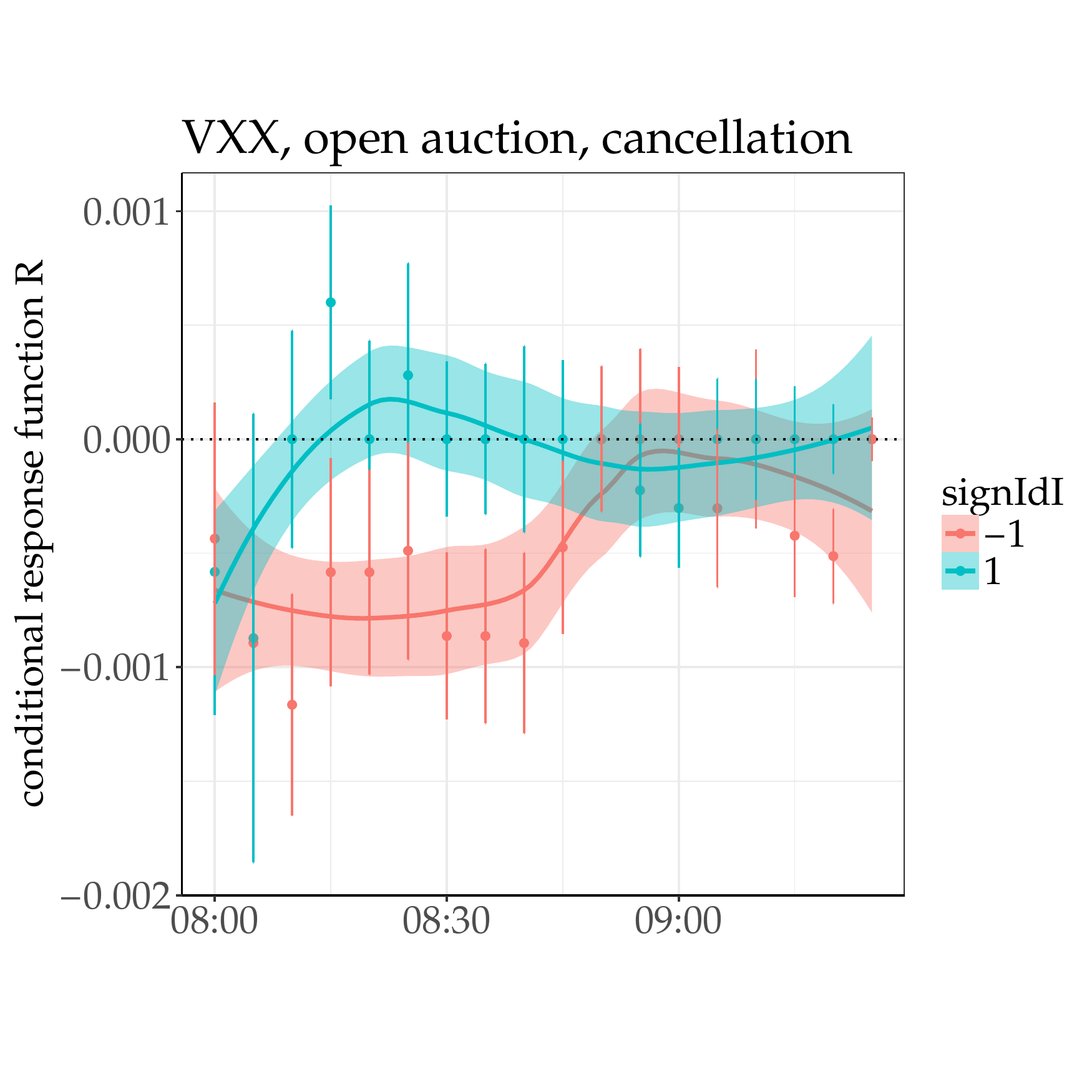}\includegraphics[width=0.5\textwidth]{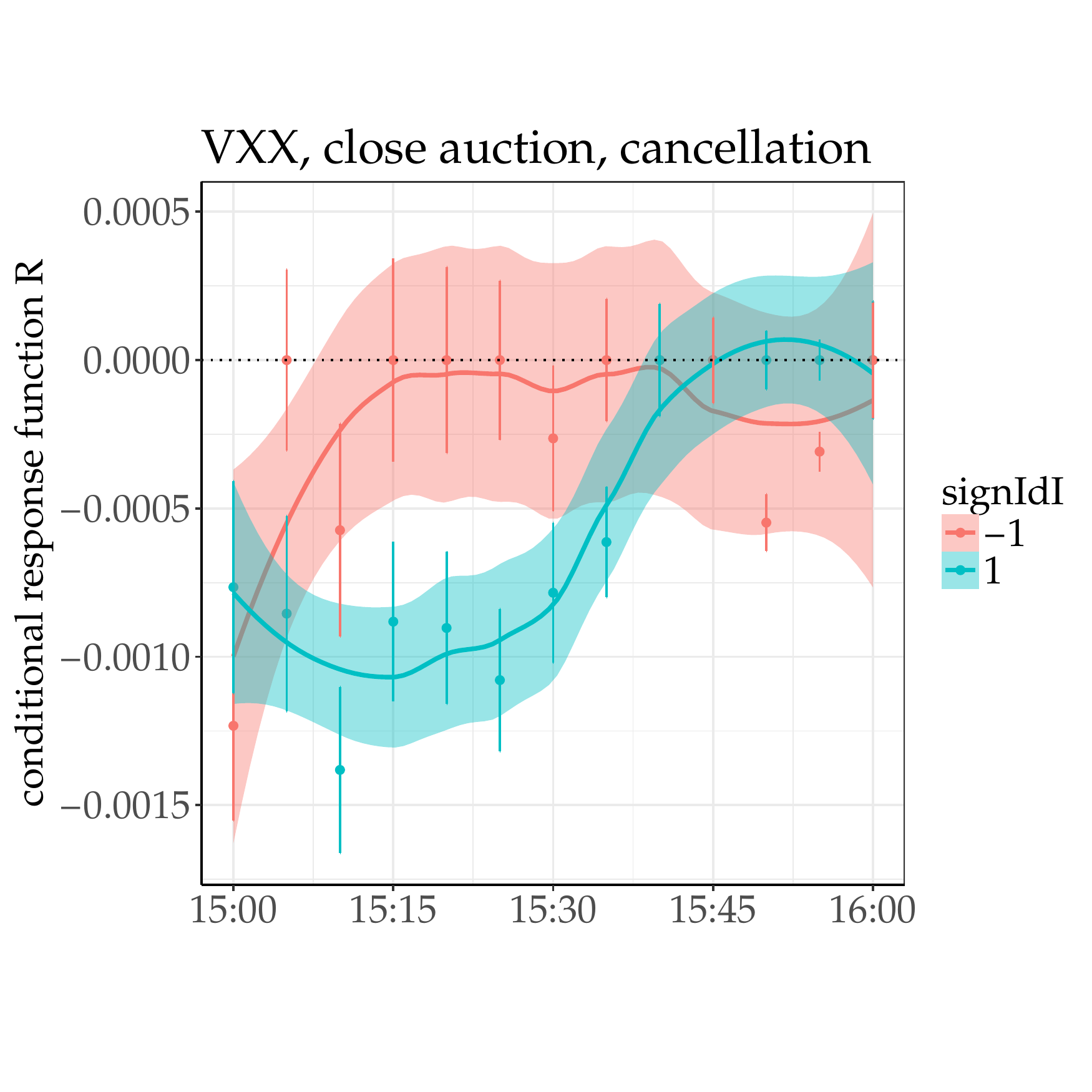}

\caption{Response functions of a new order (left plot) and order cancellation
conditional on an increase ($\text{sign }I_{t}\delta I_{t+1}=1$)
or a decrease ($\text{sign }I_{t}\delta I_{t+1}=-1$) of the absolute
total imbalance during the pre-open auctions of VXX.\label{fig:conditional_response-functions-1}
Error bars correspond to two standard deviations.}
\end{figure}

It is most revealing to discriminate between events that worsen or
improve the current imbalance. Practically, one computes a response
function for each value of $\text{sign }I(t)\times\delta I(t+1)$.
This yields a much richer picture, as show by Fig.~\ref{fig:conditional_response-functions-1}.
For many assets, there is a clear maximum of the conditional response
function of a new imbalance-worsening order. Once again, cancellations
have globally opposite effect. Most surprising
at first is the fact that for many assets, the sign of the conditional
response function reverts between the opening and the closing auctions.
This emphasizes the fundamental difference between the two auctions.
While we cannot provide a definitive explanation for this phenomenon,
we believe that it is mostly due to the fact that active trading takes
place during the closing auction, while much fewer trades take place
during the opening auction of US equities. {\color{black} Finally, note that the response functions of different assets are quite different, which reflects the fact that the typical population of traders may vary much from asset to asset. As a consequence, there is no such thing as the average response function. This is also the case for the response functions of open market limit order books, which depends for example on the exponent of the autocorrelation of market order signs \citep{bouchaudlimit4}, which in turn depends on the market participant population \citep{toth2012does}.}

{\color{black}
\subsection{Indicative price and regular limit order book}

Characterizing the dynamics of the indicative price also requires to study its interplay  with the best prices of the regular market order book. First, some mathematical notations: $a(t)$ is the best ask price, $b(t)$ the best bid price, $m(t)=[a(t)-b(t)]/2$ the mid price and $m_w(t)=\frac{v_a a-v_b b}{v_a+v_b}(t)$ the mid price weighted by the respective available volumes ($v_a$ and $v_b$) at the best prices. Intuitively, the indicative price $\pi$ must be related in some way to the mid price, since both these price are proxies of the price discovery of the same asset, although in markedly different ways. 

\begin{figure}
\centerline{\includegraphics[width=0.60\textwidth]{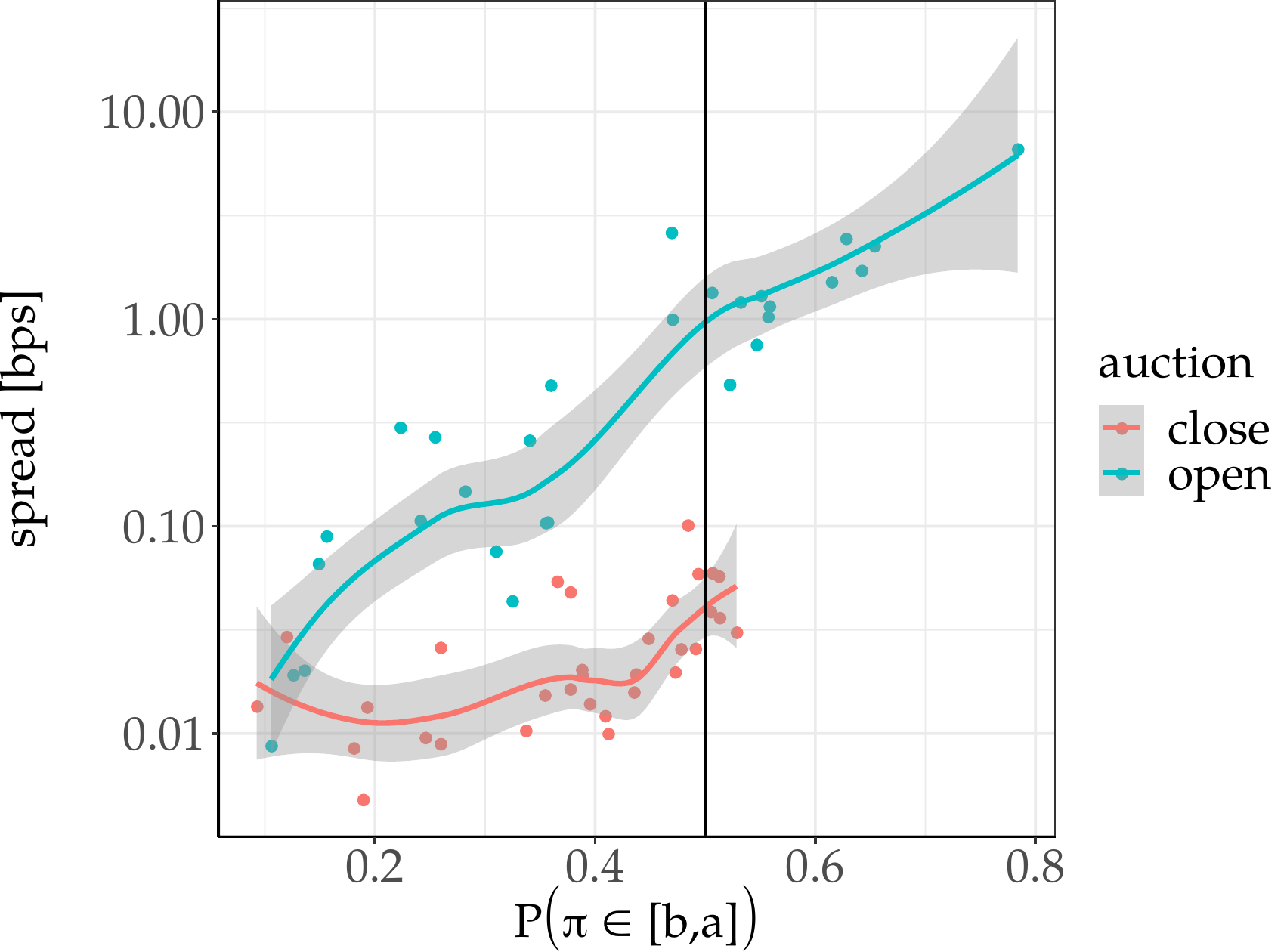}
\includegraphics[width=0.50\textwidth]{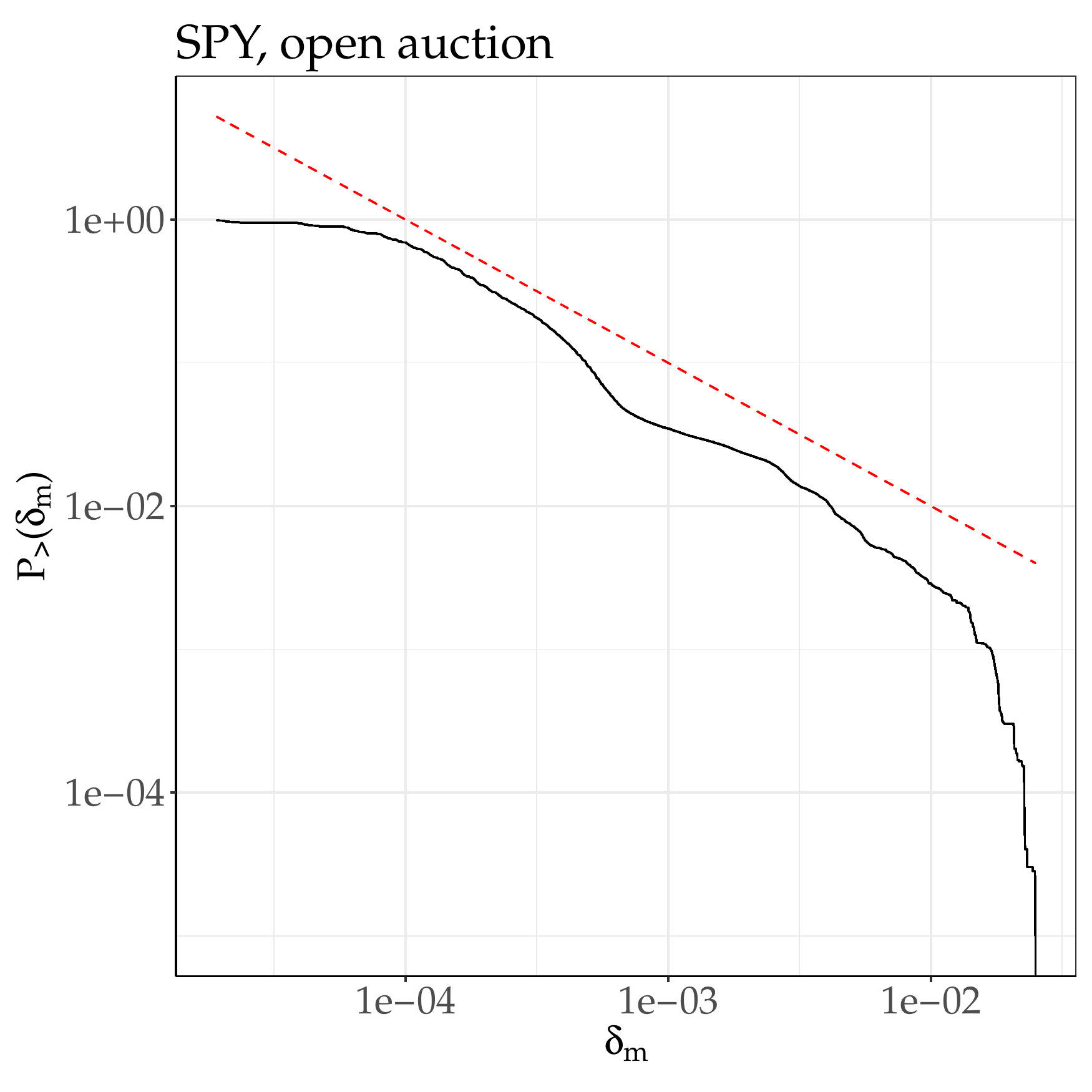}}
\caption{Left plot :median spread in bps versus the fraction of time spent in the spread by the indicative auction price for the 35 most active assets. Right plot: reciprocal cumulative distribution function $P_{>}(\delta_m=|\pi-m|/m)$ for the open auctions of SPY; the dotted red line is $\propto 1/\delta_m$ and is for eye-gudiance only.\label{fig:spread_vs_time_in_spread}}
\end{figure}

\begin{figure}
\centerline{\includegraphics[width=0.5\textwidth]{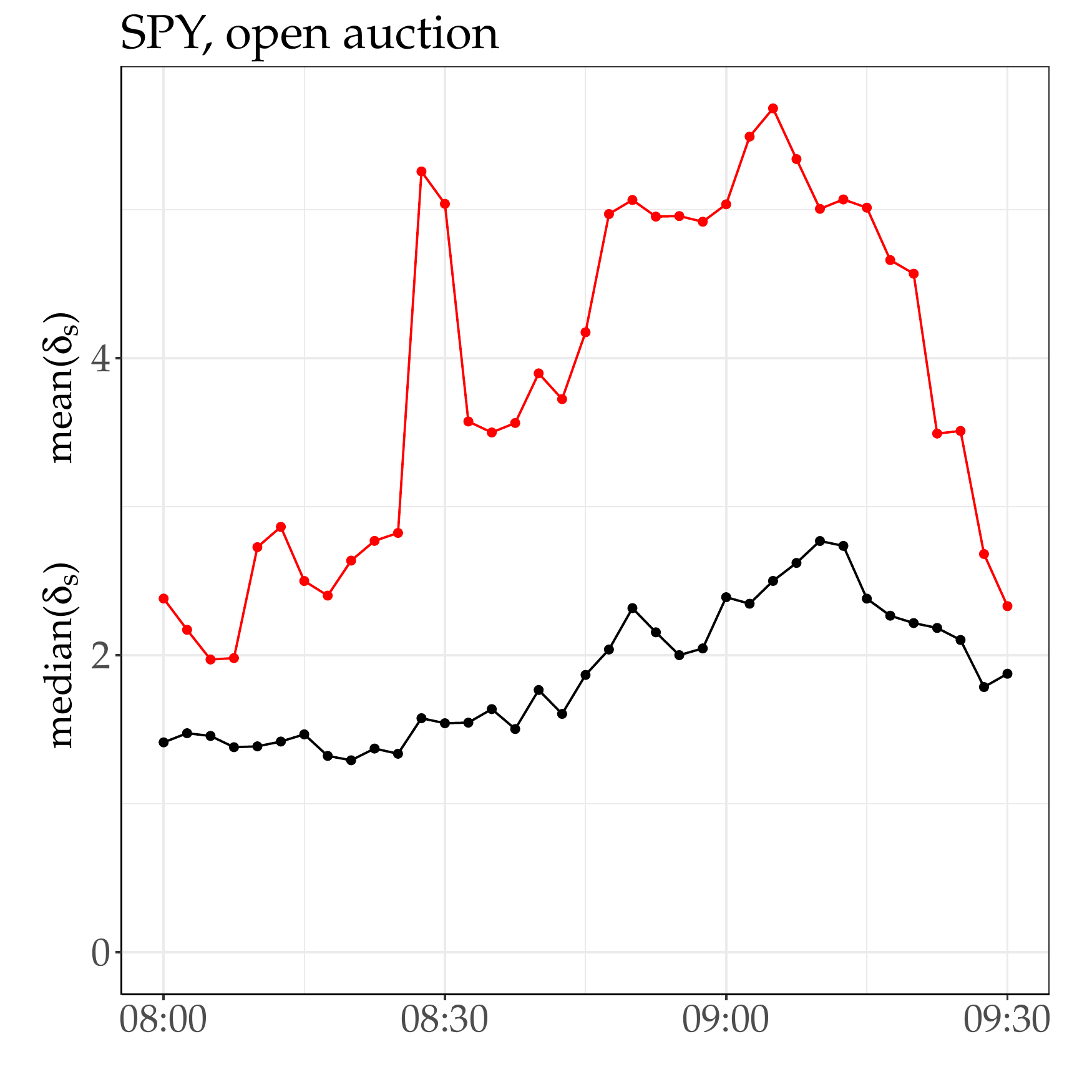}\includegraphics[width=0.5\textwidth]{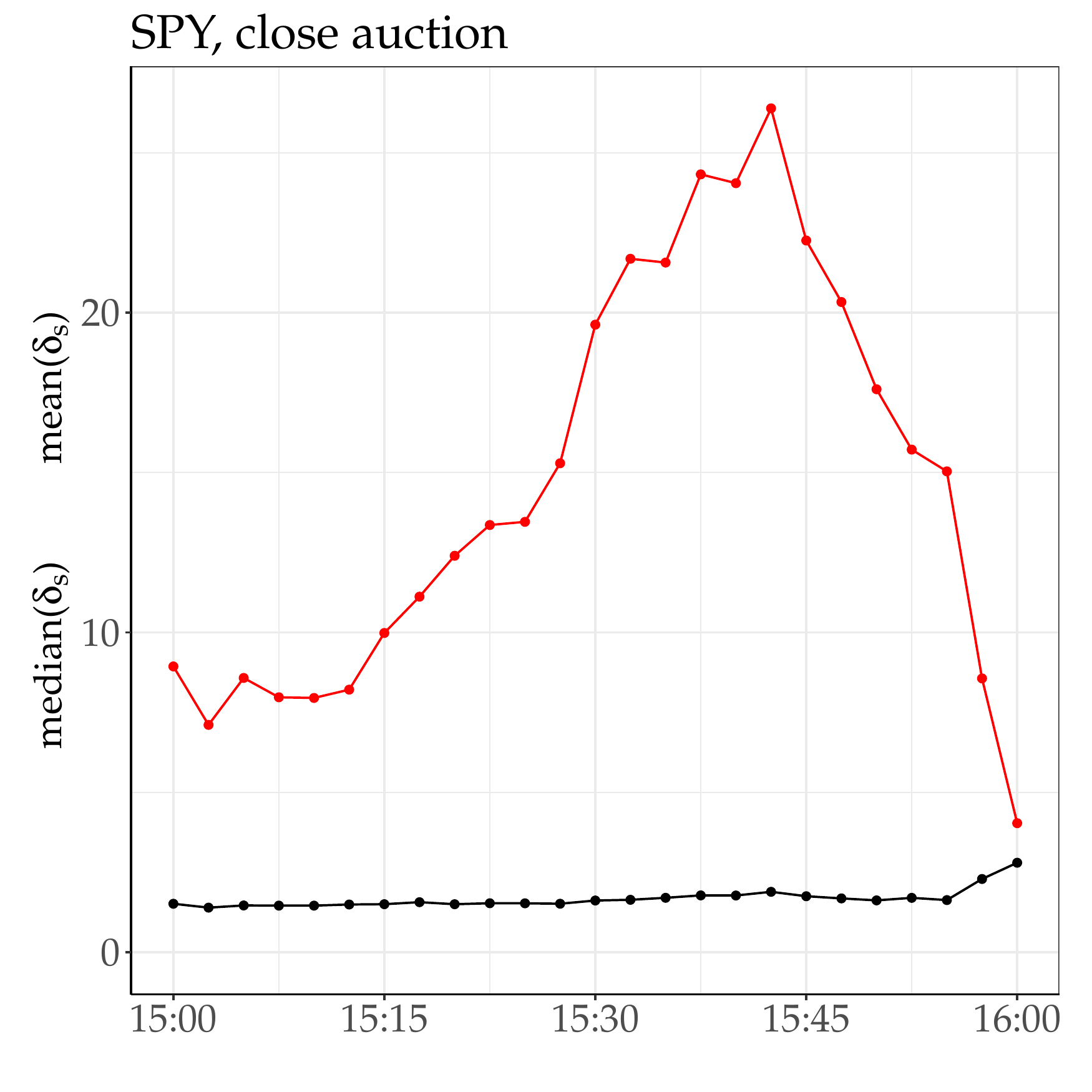}}
\caption{Mean and median distance between the indicative price and the mid price in units of spreads $\delta_s=|\pi-m|/s$ as a function time for SPY when the indicative price is outside of the spread; left plot: open auction, right plot: close auction)\label{fig:delta_s}}
\end{figure}

We first report the fraction of time (in units of updates) spent by the indicative price in the spread, i.e., $P[b(t)\le \pi(t)\le a(t)]$ for an update chosen at random in a given a time slice. Figure \ref{fig:spread_vs_time_in_spread} reports the average spread (in bps) $\tilde{s}=(a-b)/m$ versus the average $P[b(t)\le \pi(t)\le a(t)]$ for all assets (average here means average over the time-slice averages, themselves averaged over days for a given asset). For open auctions, one finds that $\tilde{s}\propto \exp P[b(t)\le \pi(t)\le a(t)]$. In other words, defining the relative deviation of the indicative price by  $\delta_m=|\pi-m|/m$ and assuming that the distribution of $\delta_m$ is the same for all these assets and independent from $\tilde{s}$, we find 

$$ P[b(t)\le \pi(t)\le a(t)]=P(\delta_m\le \tilde{s}/2)\propto \log \tilde{s},$$

hence $P(\delta_m)\propto 1/\delta_m$: this simple (and rough) argument indicates that the distribution of relative distance of the indicative price and the mid price, aggregated over all the assets, is heavy-tailed and that the indicative price cares little about being in the spread; this is due in part to the fact that the gaps between occupied ticks in the auction book are very likely heavy tailed, given the fact that regular trading hours limit order books are sparse provided that the relative tick size is not too large \citep{gillemot2006there}. We checked that there is no universal $P_>(\delta_m)$ for all the assets, but that it is generally heavy-tailed (see e.g. $P_>(\delta_m)\simeq 1/\delta_m$ for SPY in Fig.\ \ref{fig:spread_vs_time_in_spread}). Close auctions take place during regular trading hours, i.e., when the typical spread is much smaller. If the indicative price has no special reason to be within the spread, thus is not bound by a strong force to the spread, one expects that the fraction of time it spends in the spread is smaller than during the open auction. This is exactly what happens (see Fig.\ \ref{fig:spread_vs_time_in_spread}). When the indicative price is outside of the spread, it is typically less than 2 spreads away (Fig.\ \ref{fig:delta_s}), whereas its average may be much larger at times, especially when the spread is small. Interestingly, the typical mean deviation displays the same behaviour for the open and close auctions: a steady increase followed by a steeper decrease, with an additional peak at 8:30, a time at which, as remarked above, a sizeable fraction of volume is sent for this particular asset (SPY).

\begin{figure}
\centerline{\includegraphics[width=0.5\textwidth]{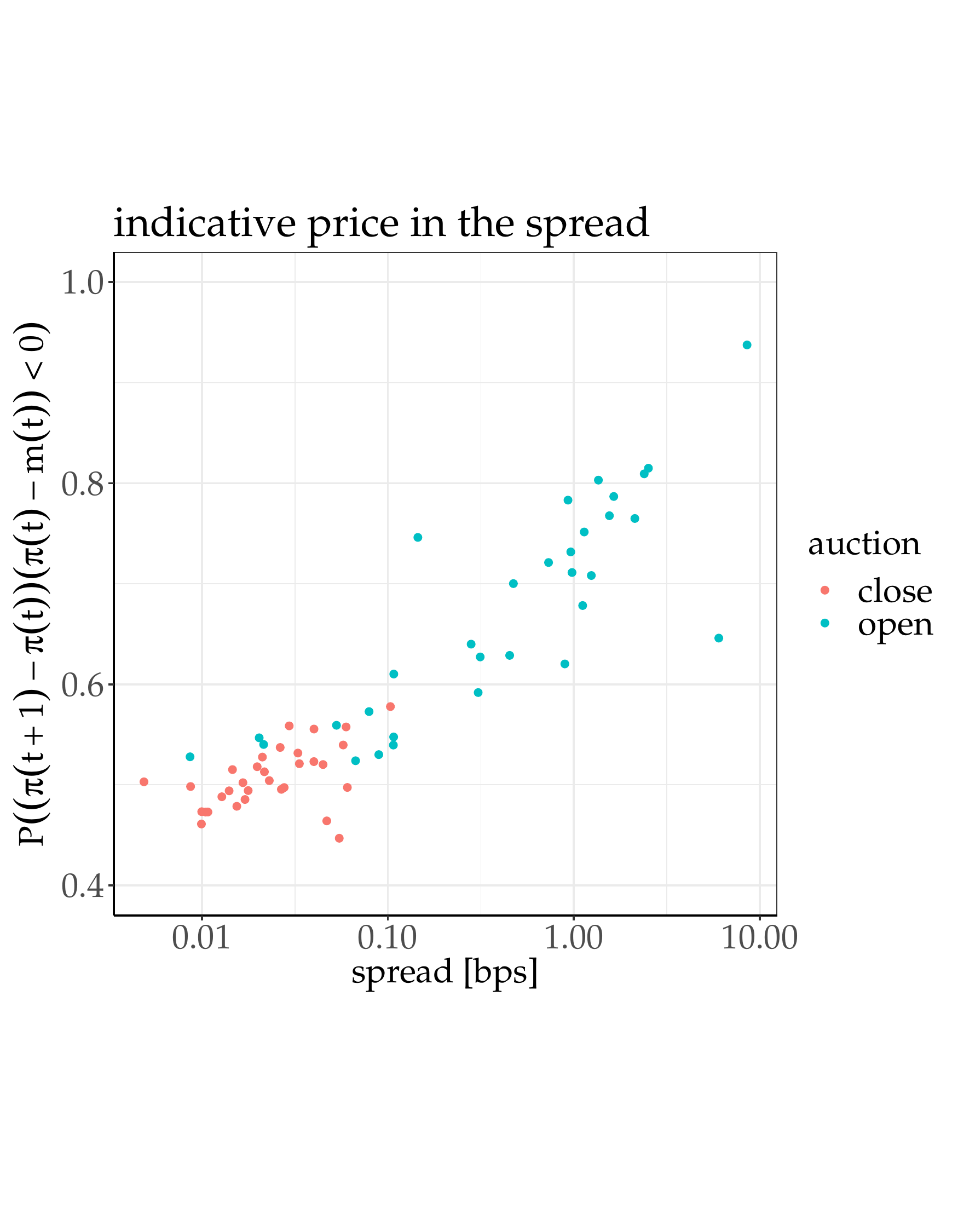}\includegraphics[width=0.5\textwidth]{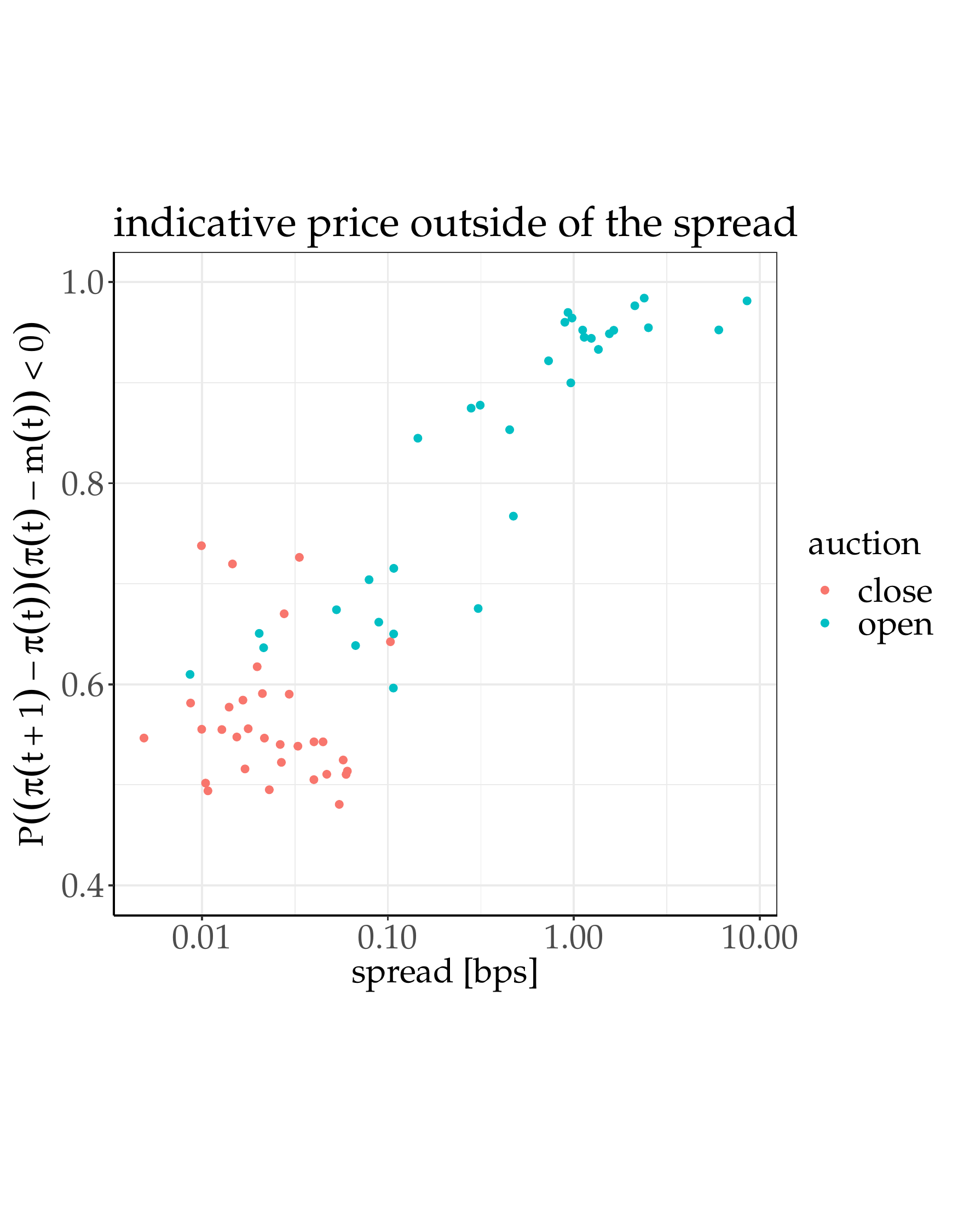}}
\caption{Probability of indicative price reversion towards the mid price as a function of the average spread when the indicative price is in the spread (left plot) and outside of the spread (right plot)\label{fig:dpipi-m}}
\end{figure}

A robust way to measure how the indicative price is attracted by the mid price is the probability that an indicative price change brings it closer to the mid price, in other words the reversion probability $P[(\pi(t+1)-\pi(t))(\pi(t)-m(t))<0]$. Figure \ref{fig:dpipi-m} plots the reversion probability  as a function of the spread; the larger the average spread, the larger the reversion probability when $\pi$ is in the spread, for both auctions. When $\pi$ is outside of the spread, the reversion does not depend on the average spread for close auctions, but clearly does for open auctions.  To check how frequent overshooting is, we also measured $P[(\pi(t+1)-m(t+1))(\pi(t)-m(t))<0|\pi \textrm{ not in spread}]$, whose average over the assets is about 0.02 for open auctions, and 0.05 for close auctions, i.e., quite small.

\begin{figure}
\centerline{\includegraphics[width=0.55\textwidth]{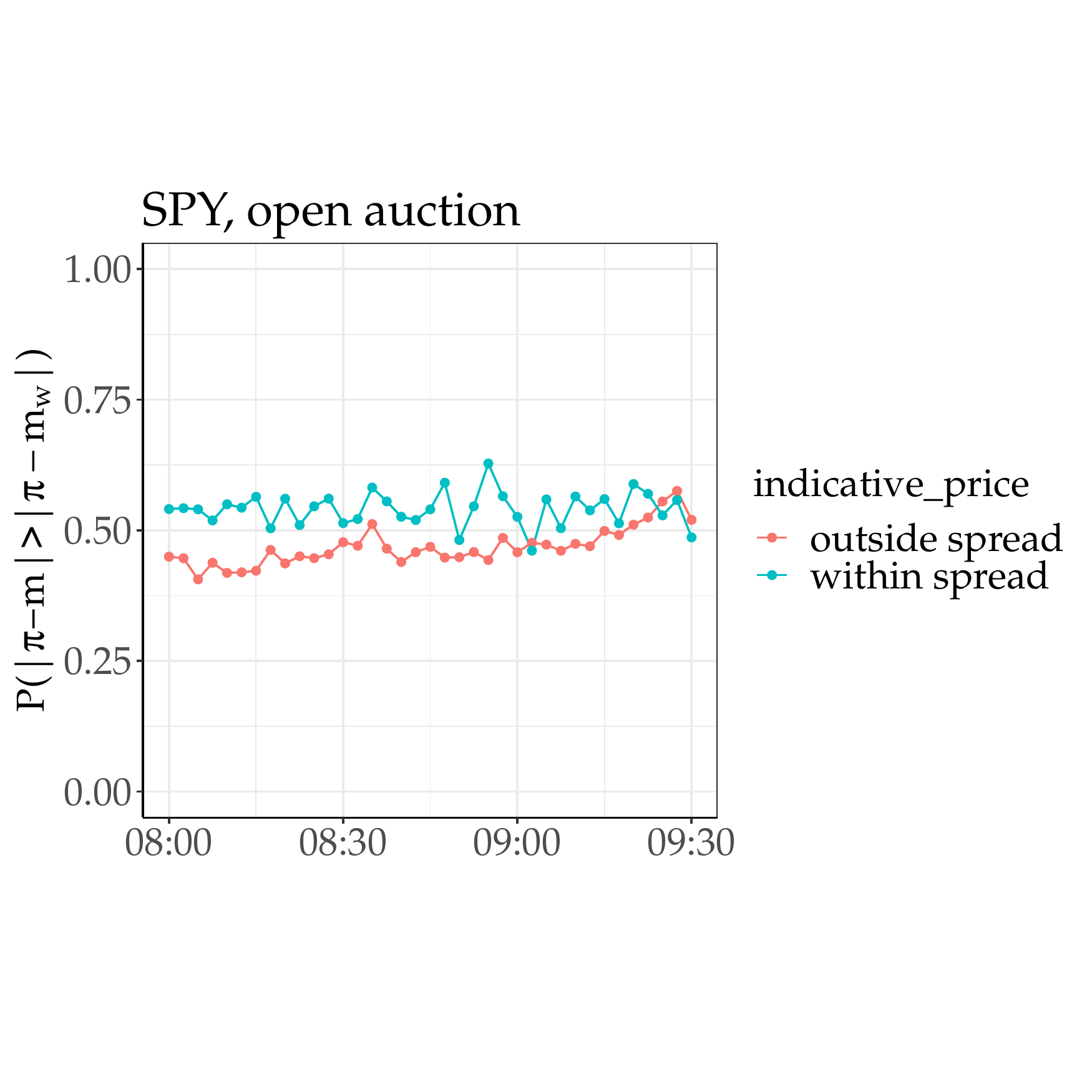}\includegraphics[width=0.55\textwidth]{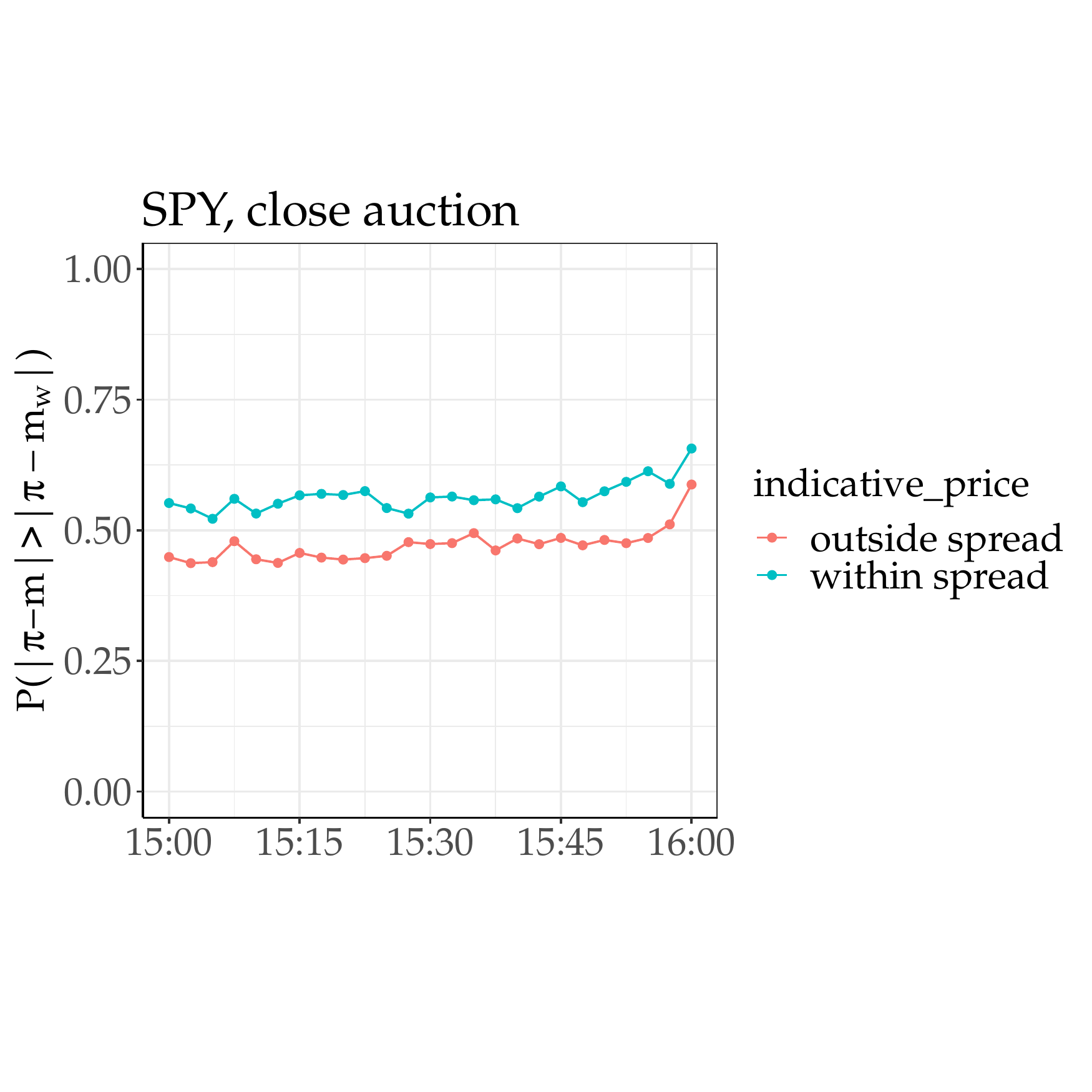}}
\caption{Probability that the indicative price is closer to the weighted mid price $m_w$ than the mid price $m$ as function of time both the open auction (left plot) and the close auction (right plot).\label{fig:mw_vs_m}}
\end{figure}

Finally, we checked if the indicative price was closer to the mid  or to the weighted mid price by estimating the fraction of times $|\pi(t)-m(t)|>|\pi(t)-m_w(t)|$ when the indicative price is in the spread and outside of it. The difference is small but real (see Fig.\ \ref{fig:mw_vs_m}). The average $P(|\pi(t)-m(t)|>|\pi(t)-m_w(t)|)\simeq 0.45$ over all time slices and the 35 most liquid assets. In other words, the mid price is a slightly more trustworthy reference point than the weighted mid price. Whether the indicative price is in or outside of the spread has no influence on the estimate.

}
\section{Conclusions}

The dynamical properties of the auction processes are far from trivial
and markedly different from those of the open-market limit order books. 
First, because the imbalance process is mean-reverting, indicative
prices are under-diffusive. Second, the median response functions
at the opening and closing auctions may have opposite signs. Three
important ingredients are responsible for these differences. First,
auctions at fixed times force the traders to reveal all liquidity
before the auction, thus increasing the importance of strategic order
submission timing, as shown at times by the large relative cancellation
of liquidity just before the auction ending time, or the acceleration of
the fraction of matched volume a few minutes before the auction  ending time
for some assets. Second, during the auction process, no trade originates
from the auction book building-up; as a consequence, usual price efficiency
conditions do not apply here. Quite notably in the US exchanges studied
here, continuous double auctions run in parallel to the auction processes;
this is one of the reasons why the opening auction response functions
are very different from those of the closing auctions: indeed, during
the closing auction, the relative liquidity available on the open-market
limit order book is much larger than that of outside regular trading
hours during the opening auction. {\color{black} Finally, the indicative price may be rather far from the mid price of the regular limit order book, although on average, the former reverts to the latter with non-negligible probability. 

Because data availability, our results on pre-auction periods are mostly about NYSE Arca, i.e., mostly about ETFs. Future work will use better data in order to check if the above findings also hold for more usual equities. Given the mechanisms discussed above, this is not unlikely.} 
 
N. G. thanks CentraleSup\'elec for an extended stay during which part
of this work was carried out. We thank Fabrizio Lillo and Thierry
Bochud for stimulating discussions. We also thank C\'edric Joulain for providing us with his LispTick fast data
streaming framework. The source code used for data
analysis is freely available at \url{https://github.com/damienchallet}.

\bibliographystyle{econ-mml}
\bibliography{biblio}

\end{document}